\documentclass[aps,pre,twocolumn,amsmath,amsfonts,showpacs]{revtex4-2}
\newcommand{\bec}[1]{\mbox{\boldmath $ #1$}}
\usepackage{bm}
\usepackage{graphicx}
\usepackage{color}

\newcommand{\meanUU}{\overline{\bm U}}
\newcommand{\meanBB}{\overline{\bm B}}
\newcommand{\meanU}{\overline{U}}
\newcommand{\meanWW}{\overline{\bm W}}

\newcommand{\meanT}{\overline{T}}
\newcommand{\meanP}{\overline{P}}
\newcommand{\meanTheta}{\overline{\Theta}}
\newcommand{\meanrho}{\overline{\rho}}

\begin{document}
\title{Large-scale semi-organized rolls in a sheared convective turbulence:\\ Mean-field simulations}
\author{A. Asulin$^{1}$}
\author{E. Tkachenko$^{1}$}
\author{N. Kleeorin$^{1,2}$}
\author{A. Levy$^{1}$}
\author{I. Rogachevskii$^{1}$}
\email{gary@bgu.ac.il}
\affiliation{
 $^1$The Pearlstone Center for Aeronautical Engineering
 Studies, Department of Mechanical Engineering,
 Ben-Gurion University of the Negev, P.O.Box 653,
 Beer-Sheva 84105,  Israel\\
 $^{2}$ Institute of Continuous Media Mechanics, Korolyov str. 1, Perm  614013, Russia}
\date{\today}
\begin{abstract}
Based on a mean-field theory of a non-rotating turbulent convection
(Phys. Rev. E {\bf 66}, 066305, 2002),
we perform mean-field simulations (MFS) of sheared convection which takes into account an effect
of modification of the turbulent heat flux by the non-uniform large-scale motions.
This effect is caused by production of additional essentially anisotropic velocity fluctuations generated by tangling of the mean-velocity gradients by small-scale turbulent motions due to the influence of the inertial forces during the lifetime of turbulent eddies.
These anisotropic velocity fluctuations contribute to the turbulent heat flux.
As the result of this effect, there is an excitation of large-scale convective-shear
instability, which causes the formation of large-scale semi-organized structures in the form of rolls.
The life-times and spatial scales of these structures are much larger compared to the turbulent scales.
By means of MFS performed for stress-free and no-slip vertical boundary conditions,
we determine the spatial and temporal characteristics of these structures.
Our study demonstrates that the modification of the turbulent heat flux by non-uniform flows
leads to a strong reduction of the critical effective Rayleigh number
(based on the eddy viscosity and turbulent temperature diffusivity)
required for the formation of the large-scale rolls.
During the nonlinear stage of the convective-shear instability,
there is a transition from the two-layer vertical structure with two rolls in the vertical direction
before the system reaches steady-state to  the one-layer vertical structure
with one roll after the system reaches steady-state.
This effect is observed for all effective Rayleigh numbers.
We find that inside the convective rolls, the spatial distribution
of the mean potential temperature includes regions with a positive vertical
gradient of the potential temperature
caused by the mean heat flux of the convective rolls.
This study might be useful for understanding of the origin
of large-scale rolls observed in atmospheric convective boundary layers
as well as in numerical simulations and laboratory experiments.
\end{abstract}

\maketitle


\section{Introduction}

Temperature stratified turbulence and turbulent convection
exist in many geophysical and astrophysical flows as well as in industrial flows
\citep[see, e.g., Refs.][]{CH61,T73,Z91,BPA00,AGL09,LX10,ZX10}.
Turbulence and associated turbulent transport were studied more than 100 years.
However, some key questions remain unclear due to extreme values of the governing parameters
in geophysical and astrophysical flows  \citep[see, e.g., Refs.][]{MY13,LE08,D13,C14,RI21}.

Large-scale coherent structures in a developed convective turbulence have been
seen in various laboratory experiments in the Rayleigh-B\'{e}nard setup \citep[see, e.g., Refs.][]{KH81,SWL89,CCL96,NSS01,NS03,BKT03,XL04,FA04,BNA05,LLT05,EEKR06,FBA08,BEKR09,BEKR11},
in the atmospheric convective turbulence  \citep[see, e.g., Refs.][]{KF84,ET85,EB93,AZ96,BR99,YKH02,W10,T17,JB21,ZK21,RKZ22,RK24},
in direct numerical simulations  \citep[see, e.g., Refs.][]{HTB03,PHP04,R06,S08,CS12,AB16} and large-eddy simulations  \citep[see, e.g., Refs.][]{HZ13,KRB17,KA19,SS20,PSS21}.
Characteristic timescales and spatial scales of the coherent structures in a
small-scale turbulent convection are much larger than the characteristic turbulent scales.

A mean-field theory of the coherent structures formed in convective turbulence
has suggested that a redistribution of the
turbulent heat flux by nonuniform large-scale motions
is crucial in the formation of the large-scale coherent
structures in a convective turbulence  (see Refs.~\cite{EKRZ02,EKRZ06,EGKR06}).
This effect causes an excitation of a convective-wind instability
in the shear-free turbulent convection resulting in the formation of
large-scale motions in the form of cells.
This phenomenon has been recently investigated by the mean-field numerical simulations
(see Ref.~\cite{OKR22}), which demonstrate that:
\begin{itemize}
\item{
The redistribution of the turbulent heat flux by the
nonuniform large-scale motions results in a strong reduction of the critical effective Rayleigh number
(based on the eddy viscosity and turbulent temperature diffusivity)
required for the formation of the large-scale convective cells.}
\item{
The convective-wind instability is excited when the scale separation ratio
between the height of the convective layer and the integral turbulence scale is large.}
\item{
The  level of the mean kinetic energy at saturation increases
with increase of the scale separation ratio, and it is very weakly dependent on the
effective Rayleigh number.}
\item{
Inside the large-scale convective cells, there
are local regions with the positive vertical gradient of the
potential temperature which implies that these regions are stably stratified.}
\end{itemize}

In the sheared convective turbulence,
the large-scale convective-shear instability results in an excitation of
convective-shear waves, and the dominant coherent structures
in the sheared convection are rolls (see Refs.~\cite{EKRZ02,EKRZ06,EGKR06}).
The goal of the present study is to perform mean-field numerical simulations
in a sheared convection taking into account the effect of modification of the
turbulent heat flux by non-uniform large-scale motions.

This paper is organized as follows.
In Section~II we discuss physics related to
a modification of the turbulent heat flux due to anisotropic velocity fluctuations in turbulence
with non-uniform large-scale flows in a sheared convective turbulence.
In Section~III we formulate the non-dimensional equations, the governing non-dimensional parameters
and study the large-scale convective-shear instability.
In Section~IV we describe the set-up for the mean-field simulations
and discuss the  numerical results.
Finally, conclusions are drawn in Section~V.

\section{Sheared turbulent convection and turbulent heat flux}

We consider sheared turbulent convection with very high Rayleigh numbers, and large
Reynolds and Peclet numbers.
To study formation and evolution of the semi-organized structures
in a small-scale convective turbulence, we use a mean field approach.
In the framework of this approach, the velocity ${\bm U}$, pressure $P$ and
potential temperature $\Theta$ are decomposed into the mean fields and fluctuations, where
${\bm U} = \meanUU + {\bm u}$, $P = \meanP + p$ and $\Theta =
\meanTheta + \theta$.
Since we use Reynolds averaging, fluctuations have zero mean values,
where $\meanUU = \langle {\bm U} \rangle$ is the mean velocity, $\meanP = \langle P
\rangle$ is the mean pressure and  $\meanTheta = \langle \Theta \rangle$
is the mean potential temperature, and ${\bm u}$, $p$ and $\theta$
are fluctuations of velocity, pressure and potential temperature, respectively.
Averaging the Navier-Stokes equation and equation for the potential temperature
over an ensemble, we arrive at the mean-field equations written in the Boussinesq approximation with ${\rm div} \, \meanUU= 0$:
\begin{eqnarray}
&& \biggl[{\partial  \over \partial t} + \left(\meanUU + \meanUU_{\rm S}\right) \cdot {\bm \nabla}\biggr] \meanUU
+ (\meanUU \cdot {\bm \nabla}) \meanUU_{\rm S}
\nonumber \\
&& \quad \quad \quad = - {{\bm \nabla} \meanP \over \meanrho_0}
- {\bm \nabla} \cdot \hat{\bm R} + \beta \, \meanTheta \, {\bm e}_z,
\label{LB1}
\end{eqnarray}
\begin{eqnarray}
&&\biggl[{\partial  \over \partial t} + \left(\meanUU + \meanUU_{\rm S}\right) \cdot {\bm \nabla}\biggr] \meanTheta
= - (\meanUU \cdot {\bm \nabla}) \meanT_0
- {\bm \nabla} \cdot  \langle {\bm u} \theta \rangle ,
\nonumber \\
\label{LB2}
\end{eqnarray}
where $\hat{\bm R} \equiv R_{ij}  = \langle u_{i} \, u_{j} \rangle$ is the Reynolds stress,
$\meanUU_{\rm S}=(S \,z,0,0)$ is the shear velocity directed along the $x$ axis,
$\meanTheta$ is the mean potential temperature defined as
$\meanTheta=\meanT \,(\meanP_0/\meanP)^{1-1/\gamma}$.
Here $\beta = |{\bm g}|/T_0$  is the buoyancy parameter,
${\bm g}=-g \, {\bm e}_z$ is the acceleration caused by the gravity,
${\bm e}_z$ is the unit vector in the vertical direction  (along the $z$ axis),
$\gamma=c_{\rm p}/c_{\rm v}$ is the specific heats ratio,
$\meanT$ is the mean physical temperature with the reference value
$\meanT_0$ as the temperature in the equilibrium (i.e.,  the basic reference state),
$\meanP$ is the mean pressure with the reference value $\meanP_0$
and $\meanrho_0$ is the mean fluid density in the equilibrium.
For large Reynolds and Peclet numbers, we neglect
in Eqs.~(\ref{LB1})--(\ref{LB2}) small terms due to the kinematic viscosity
and molecular diffusivity of the potential temperature in comparison with
those due to the turbulent viscosity and turbulent diffusivity.
In Eqs.~(\ref{LB1})--(\ref{LB2}), the mean fields
corresponds to deviations from the equilibrium: $\meanUU_{0}=\meanUU_{\rm S}$, $\; {\bm \nabla} \meanP_{0} = \meanrho_0 {\bm g}$ and $\meanrho_0 =$ const.

The effects of small-scale convective turbulence on the mean fields are described
by the Reynolds stress $\langle u_{i} \, u_{j} \rangle$ and turbulent flux of potential temperature
${\bm F} = \langle {\bm u} \theta \rangle$.
In the classical concept of down-gradient turbulent transport,
the basic second-order moments (e.g., the Reynolds stress
and the turbulent flux of potential temperature) are assumed to be proportional
to the local mean gradients, whereas the proportionality
coefficients, namely turbulent viscosity $\nu_{_{T}}$
and turbulent temperature diffusivity $\kappa_{_{T}}$, are
determined by local turbulent parameters.
For instance, the Reynolds stress is
$\langle u_{i} \, u_{j} \rangle = - 2 \nu_{_{T}} (\nabla_i \meanU_j + \nabla_j \meanU_i)$, while
the turbulent heat flux is given by
${\bf F} = - \kappa_{_{T}} \bec{\nabla} \meanTheta$ \cite{MY13}.

In turbulent convection with semi-organized structures
(e.g., large-scale circulations and large-scale convective rolls),
the mean velocity and temperature fields inside the semi-organized structures
are strongly nonuniform.
These nonuniform large-scale motions produce strongly
anisotropic velocity fluctuations which contribute
to the turbulent heat flux.
As has been shown in Refs.~\cite{EKRZ02,EKRZ06},
the turbulent heat flux ${\bf F}$ which takes into account
anisotropic velocity fluctuations, reads
\begin{eqnarray}
{\bf F} &=& {\bf F}^{\ast} -  \tau_0 \, \biggl[{\bf
F}_{z}^{\ast} \, {\rm div} \, \meanUU_{\perp} - {1 \over 2} \,
\left(\meanWW {\bf \times} {\bf F}_z^{\ast}\right)
\nonumber\\
&& - {1 \over 2} \, \left(\meanWW_z {\bf \times} {\bf F}_x^{\rm CW}\right)
\biggr] ,
 \label{R100}
\end{eqnarray}
where ${\bf F}_x^{\rm CW}= - \tau_0 \,({\bf F}_z^{\ast} \cdot \bec{\nabla})\meanUU_{\rm S}(z)$
is the counter-wind turbulent heat flux
and ${\bf F}^{\ast} = - \kappa_{_{T}} \bec{\nabla} \meanTheta$
is the classical turbulent heat flux,
$\tau_0$ is the correlation time of turbulent velocity
at the integral scale of turbulent motions, $\meanWW =
\bec{\nabla} {\bf \times}  \meanUU$ is the mean vorticity,
$\meanUU=  \meanUU_{\perp} +  \meanUU_{z}$ is the mean velocity with the horizontal
$\meanUU_{\perp}$ and vertical $\meanUU_{z}$ components.
Equation~(\ref{R100}) is derived for large Reynolds and Peclet numbers,
which implies that turbulent viscosity and turbulent diffusivity
are much larger than the molecular viscosity and molecular diffusivity, respectively.
The additional terms  in the turbulent heat flux result in
the excitation of large-scale instability and formation of
the large-scale convective rolls \cite{EKRZ02,EKRZ06,EGKR06}.

\begin{figure}
\vspace*{1mm}
\centering
\includegraphics[width=8cm]{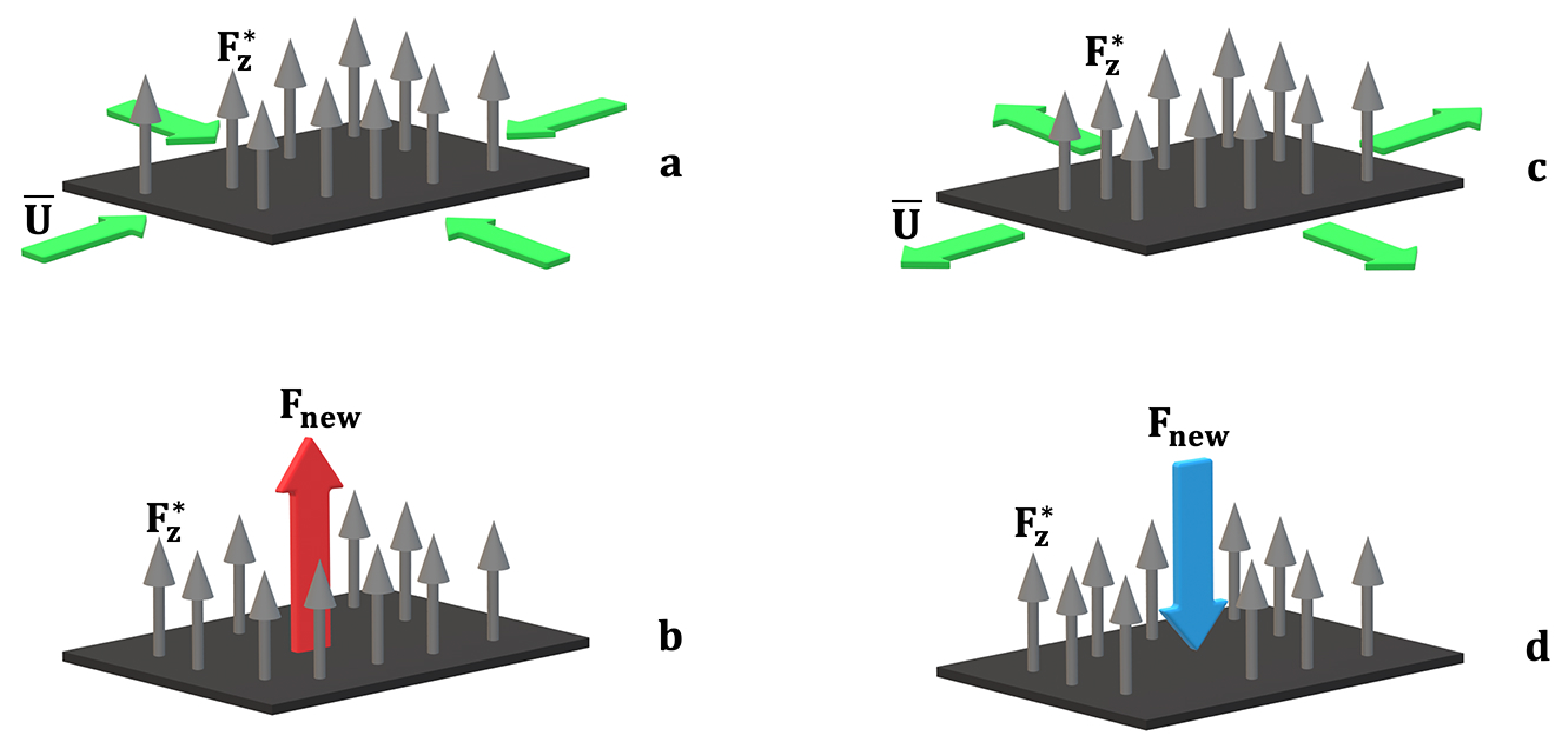}
\caption{\label{Fig1} The illustration of the physics caused by the new turbulent heat flux
${\bm F}_{\rm new} = - \tau_0 \, {\bm F}_{z}^{\ast} \, {\rm div}
\, \meanUU_{\perp}$ produced by the perturbations of the
convergent (or divergent) horizontal mean flows $\meanUU_{\perp}$ (shown by the green arrows
in panels {\bf a} and {\bf c}).
The new turbulent flux ${\bm F}_{\rm new}$ increases the upward turbulent heat flux, enhances buoyancy and increases the local mean potential temperature, thus creating the upward flow. Likewise, the new turbulent flux ${\bm F}_{\rm new}$ decreases the vertical turbulent flux of potential temperature by the divergent horizontal motions, which reduces the buoyancy and decreases the local mean potential temperature, thus creating the downward flow. These effects create the large-scale circulation.
}
\end{figure}

The physics related to the additional terms in the turbulent heat flux is discussed below.
The term $ - \tau_0 \, {\bm F}_{z}^{\ast} \, {\rm div}
\, \meanUU_{\perp}$ in Eq.~(\ref{R100}) for the turbulent heat flux
causes the redistribution of the vertical background turbulent
heat flux ${\bm F}_{z}^{\ast}$ by the perturbations of the
convergent (or divergent) horizontal mean velocity $\meanUU_{\perp}$ (see Fig.~\ref{Fig1})
during the life-time of turbulent eddies.
This enhances the vertical turbulent flux
of potential temperature due to the converging horizontal motions, which
increases the buoyancy, thus creating the upward flow.
The latter increases the horizontal convergent flow.

On the other hand, the term $ (\tau_0/2) \, (\meanWW {\bm \times}
{\bm F}_z^{\ast})$ in Eq.~(\ref{R100}) produces
the horizontal turbulent heat flux by the
"rotation" of the vertical background turbulent heat flux ${\bm
F}_{z}^{\ast}$  caused by the perturbations of the horizontal mean
vorticity $\meanWW_{\perp}$.
This decreases local potential temperatures
in rising motions, which decreases the buoyancy accelerations,
and weakens vertical velocity and vorticity.

The last term $ (\tau_0 /2) \, \left(\meanWW_z {\bf \times} {\bf F}_x^{\rm CW}\right)$ in Eq.~(\ref{R100})
for the turbulent heat flux produces the horizontal heat flux through the ''rotation"
of the horizontal background counter-wind turbulent heat flux ${\bf F}_x^{\rm CW}$ by the
vertical component of the mean vorticity.
The counter-wind turbulent flux of potential temperature
${\bf F}_x^{\rm CW}$ arises  due to the following reasons.
In a horizontally homogeneous and sheared convective turbulence,
the mean shear velocity $\meanU^{\rm S}_x(z)$ increases with the height,
while the mean potential temperature $\overline{\Theta}(z)$ decreases with the height.
Uprising fluid particles produce both, positive fluctuations of potential temperature, $\theta>0$
because $\partial \theta/\partial t \sim - ({\bm u} \cdot
\bec{\nabla}) \overline{\Theta}(z)$, and negative fluctuations of horizontal velocity, $u_x<0$
because $\partial u_x/\partial t \sim - ({\bm u} \cdot
\bec{\nabla}) \meanU^{\rm S}_x(z)$.
This creates a negative horizontal turbulent flux of potential temperature, $u_x \, \theta<0$.
On the other hand, sinking fluid particles create both, negative fluctuations of potential temperature, $\theta<0$,
and positive fluctuations of horizontal velocity, $u_x>0$, resulting in
negative horizontal turbulent flux of potential temperature,
$u_x \, \theta<0$. Therefore, the net horizontal turbulent flux of potential temperature is negative,
${\bf F}_x^{\rm CW} \equiv \langle u_x \, \theta\rangle<0$,
in spite of a zero horizontal mean temperature gradient.
Therefore, the counter-wind turbulent flux of potential temperature
modifies the turbulent potential temperature flux caused by
the non-uniform mean velocity field.
The counter-wind turbulent flux is associated with non-gradient turbulence transport of heat.

\begin{figure}
\vspace*{1mm}
\centering
\includegraphics[width=8cm]{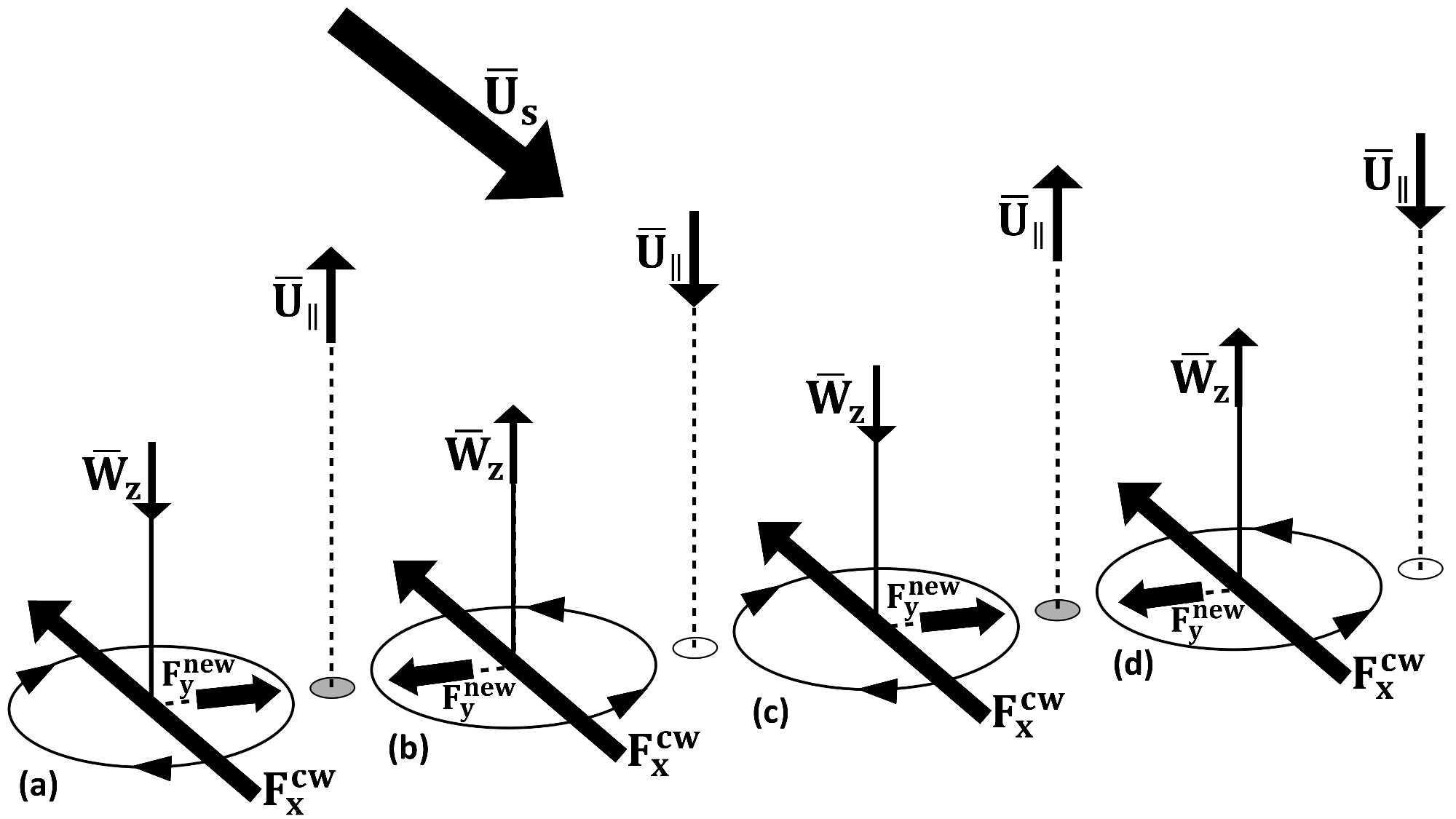}
\caption{\label{Fig2} The mechanism of formation of large-scale convective rolls
stretched along the shear velocity $\meanUU_{\rm S}(z)$. Horizontal counter-wind turbulent heat flux ${\bf F}_x^{\rm CW}$ is turned by perturbations of vertical vorticity $\meanWW_z$, i.e., this effect creates alternating pairs of convergence or divergence cross-wind turbulent heat fluxes, ${\bf F}_y^{\rm new}=(\tau_0/2) \, \meanWW_z {\bf \times} {\bf F}_x^{\rm CW}$, which are clock-wise or opposite rotations of air columns.
This effect causes converging turbulent heat fluxes with warm patch between
the pair  (a) and (b) of columns in the Figure, and diverging turbulent heat fluxes
with cool patch between the pair (b) and (c) in the Figure.
The warm patch causes an updraft and cool patch produces a downdraft,
whereas the mean shear velocity $\meanUU_{\rm S}(z)$ stretches the flow pattern
and completes creation of the large-scale convective rolls.}
\end{figure}

\begin{figure}
\vspace*{1mm}
\centering
\includegraphics[width=8cm]{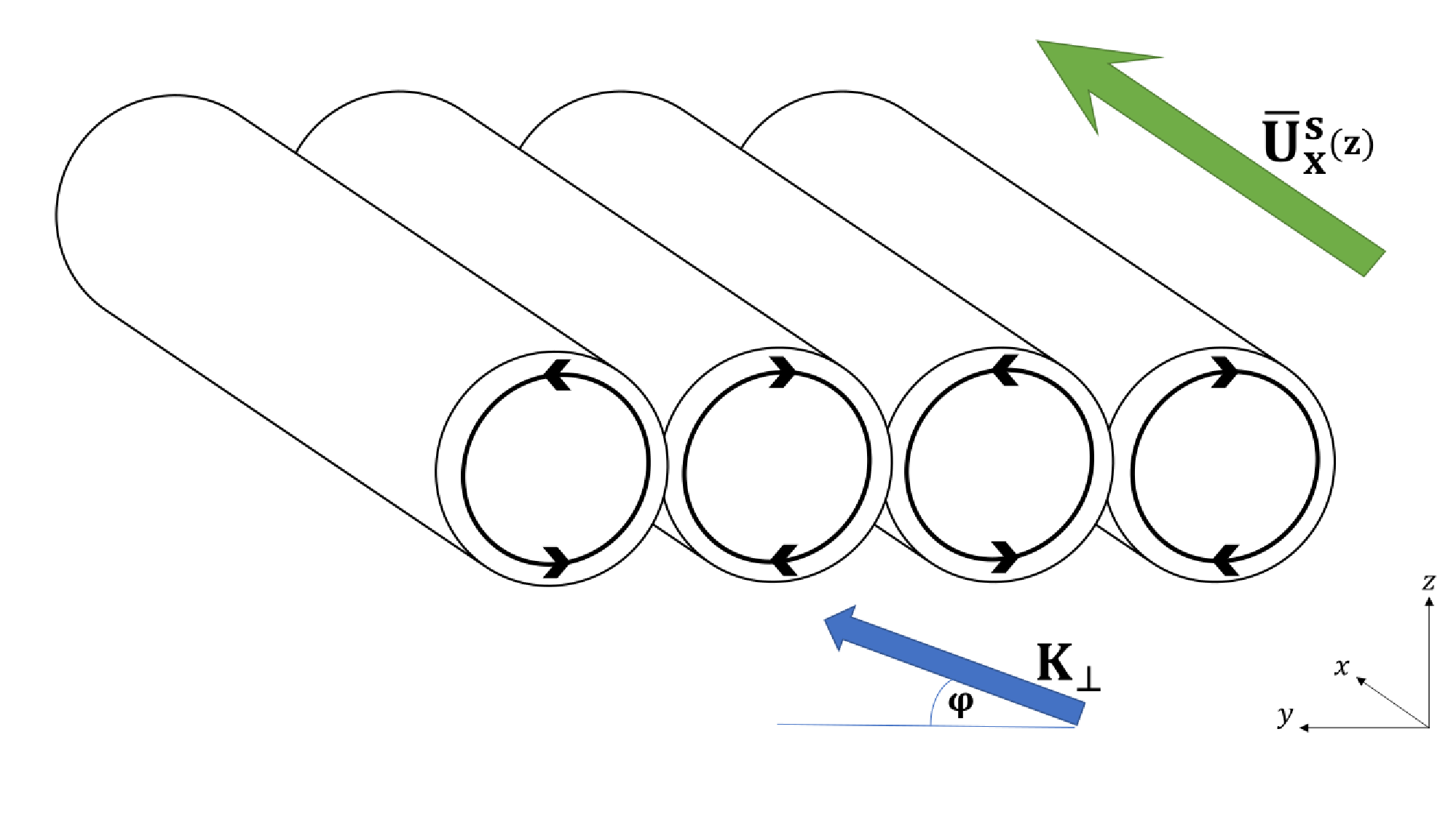}
\caption{\label{Fig3}
Large-scale convective rolls stretched along the mean velocity shear and generation of convective-shear waves propagating perpendicular to the convective rolls.}
\end{figure}

The last term $(\tau_0/2) \, \meanWW_z {\bf \times} {\bf F}_x^{\rm CW}$ in Eq.~(\ref{R100}),
causes generation of the cross-wind horizontal turbulent heat flux by turning the counter-wind horizontal turbulent
flux  ${\bf F}_x^{\rm CW}$ by perturbations of vertical component of the mean vorticity $\meanWW_z$ (see Fig.~\ref{Fig2}).
This produces alternating pairs of convergence or divergence cross-wind turbulent heat fluxes, ${\bf F}_y^{\rm new}=(\tau_0/2) \, \meanWW_z {\bf \times} {\bf F}_x^{\rm CW}$, resulting alternative warmer or cooler patches which, in turn, cause alternative upward warm and downward cool motions. This is precisely the mechanism of large-scale instability responsible for formation of the large-scale convective rolls stretched along the mean velocity shear (see Fig.~\ref{Fig3}) and generation of convective-shear waves propagating perpendicular to the convective rolls
in the sheared convection \cite{EKRZ02,EKRZ06,EGKR06}.

\section{Governing equations and convective-shear instability}

Using the expression~(\ref{R100}) for the turbulent heat flux ${\bf F}$ with the additional terms
caused by the non-uniform mean flows,
calculating div ${\bm F}$, and assuming that the non-dimensional total vertical heat flux
$\Phi_c=\tilde F_z^\ast + \tilde U_z \, \tilde\Theta$ is constant,
we rewrite Eqs.~(\ref{LB1})--(\ref{LB2}) in a non-dimensional form as
\begin{eqnarray}
&& {\partial \tilde{\bm U} \over \partial t} + \left[\left(\tilde{\bm U} + \tilde{\bm U}_{\rm S}\right) \cdot {\bm \nabla}\right]
\tilde{\bm U} + (\tilde{\bm U} \cdot {\bm \nabla}) \tilde{\bm U}_{\rm S}
\nonumber\\
&& \;
= - {{\bm \nabla} \tilde P \over \rho_0} + {\rm Ra}_{_{T}} \, \tilde\Theta \, {\bm e}_z + \Delta \tilde{\bm U},
 \label{B1}
\end{eqnarray}

\begin{eqnarray}
&&  {\rm Pr}_{_{T}} \, \left({\partial \tilde\Theta \over \partial t} + \left[\left(\tilde{\bm U} + \tilde{\bm U}_{\rm S}\right) \cdot {\bm \nabla}\right] \tilde\Theta\right) =  \tilde U_z +  \Delta \tilde\Theta
\nonumber\\
&& \;
+ {\epsilon \over 2} \biggl\{ \left(\Phi_c - \tilde U_z \, \tilde\Theta \right) \, \biggl[\left(\Delta - 2\nabla_z^2\right) \tilde U_z + {\rm Sh}  \, \Big(\nabla_y^2 \tilde U_x
\nonumber\\
&& \; - \nabla_x \nabla_y \tilde U_y \Big)\biggr]
+ 2 \left(\nabla_z \tilde U_z\right)
\nabla_z \left( \tilde U_z \, \tilde\Theta \right)
+ \Big(\nabla_z \tilde U_x
\nonumber\\
&& \;
- \nabla_x \tilde U_z\Big) \, \nabla_x \left(\tilde U_z \, \tilde\Theta \right)
+ \biggl[\nabla_z \tilde U_y - \nabla_y \tilde U_z
\nonumber\\
&& \;
+ {\rm Sh} \,\left(\nabla_x \tilde U_y - \nabla_y \tilde U_x\right)\biggr]
\, \nabla_y \left( \tilde U_z \, \tilde\Theta \right)
\biggr\},
  \label{B2}
\end{eqnarray}
where the non-dimensional mean velocity $\tilde{\bm U}$ with ${\rm div} \, \tilde{\bm U}= 0$, the mean potential temperature $\tilde\Theta$ and the mean pressure $\tilde P$ are shown with tilde, the flux $\tilde F_z^\ast$ is the nondimensional vertical turbulent
background heat flux,
the unit vector ${\bm e}_z$ is directed along vertical $z$ axis,
the non-dimensional shear velocity is $\tilde{\bm U}_{\rm S} = ({\rm Sh}\, \tilde z, 0, 0)$,
and $\tilde z$ is is the non-dimensional vertical coordinate.

Equations~(\ref{B1})--(\ref{B2}) are written in non-dimensional form, where
length is measured in the units of the vertical size of the convective layer $L_z$,
time is measured in the units of
the turbulent viscosity time, $L_z^2/\nu_{_{T}}$, velocity is measured in the units
of $\nu_{_{T}}/L_z$, potential temperature is measured in the units of $L_z \, N^2 \, {\rm Pr}_{_{T}}/\beta$
and pressure is measured in the units of $\rho_0 \, (\nu_{_{T}}/L_z)^2$.
Here $\nu_{_{T}}=u_0 \, \ell_0/3$ is the turbulent (eddy) viscosity, $u_0$ is the r.m.s. turbulent velocity,
$\ell_0$ is the turbulent integral scale and $N^2 = \beta \, |\nabla_z \meanT_{\rm eq}|$.

We use the following dimensionless parameters appeared in Eqs.~(\ref{B1})--(\ref{B2}):
\begin{itemize}
\item{
 the effective Rayleigh number:
\begin{eqnarray}
{\rm Ra}_{_{T}} = {L_z^4 \, N^2 \over \nu_{_{T}} \, \kappa_{_{T}}} ,
\label{B3}
\end{eqnarray}
}
\item{
 the turbulent Prandtl number:
\begin{eqnarray}
{\rm Pr}_{_{T}} = {\nu_{_{T}} \over \kappa_{_{T}}} ,
\label{B4}
\end{eqnarray}
}
\item{
 the scale separation parameter:
\begin{eqnarray}
\epsilon = {\ell_{0}^2 \over 3 L_z^2} ,
\label{B5}
\end{eqnarray}
}
\item{
 the non-dimensional total vertical heat flux:
\begin{eqnarray}
\Phi_c = {3 \over \epsilon^2\, {\rm Ra}_{_{T}}} \, \left({u_c \over u_{0}}\right)^3  ,
\label{B6}
\end{eqnarray}
}
\item{
 the non-dimensional shear number:
\begin{eqnarray}
{\rm Sh} = {S \, \tau_0 \over \epsilon} ,
\label{BS6}
\end{eqnarray}
}
\end{itemize}
where $S = \nabla_z \meanU^{\rm S}_x$ is the linear velocity shear, i.e.,
$S$ is constant, $u_c = (g F_z \ell_{0})^{1/3}$ is the convective velocity,
$F_z$ is the vertical turbulent flux of potential temperature,
$\kappa_{_{T}}$ is the turbulent (eddy) diffusivity
and $\tau_0=\ell_0/u_0$.
In Eqs~(\ref{B1})--(\ref{B2}) we have neglected small terms $\sim {\rm O}(\epsilon^2)$.

For analytical study of the large-scale instability, we consider for simplicity
the two-dimensional problem when the mean fields are independent
of the coordinate $x$.
The non-dimensional shear velocity, $\tilde{\bm U}_{\rm S} = ({\rm Sh}\, \tilde z, 0, 0)$,
is directed along the $x$ axis, so that the vorticity $\tilde{\bm W}_{\rm S} \equiv
{\bm \nabla} \times \tilde{\bm U}_{\rm S}$
is $\tilde{\bm W}_{\rm S} = (0, {\rm Sh}, 0)$.
We will start our analysis with the linear problem for small perturbations applying the linearised Eqs.~(\ref{B1})--(\ref{B2}),
to find the growth rate of the large-scale convective-shear instability.
To this end, we calculate $[{\bm \nabla} \times ({\bm \nabla} \times \tilde{\bm U})]_z$
using the linearised  Eq.~(\ref{B1}) to exclude the pressure term and we
seek for solution of the obtained equations in the following form:
$\tilde {\bm U}(t,{\bm x}) = \tilde {\bm U}_0 \exp[\gamma t - {\rm i} (K_y y + K_z z)]$
and $\tilde \Theta(t,{\bm x}) = \Theta_0 \exp[\gamma t - {\rm i} (K_y y + K_z z)]$.
This yields the following system of algebraic equations:
\begin{eqnarray}
&& \left(\gamma + K^2\right)  \,\tilde U_z + {\rm Ra}_{_{T}} \, \left({K_z^2 \over K^2} -1\right) \, \tilde \Theta =0 ,
\label{AAA3}\\
&& \left[1 + {\sigma  \over 2\epsilon \, {\rm Ra}_{_{T}}} \left(2K_z^2 - K^2  + {{\rm Sh}^2 K_y^2 \over \gamma + K^2} \right) \right] \,\tilde U_z
\nonumber\\
&& \qquad - \left(\gamma + K^2\right) \,\tilde \Theta =0 ,
\label{AAA4}
\end{eqnarray}
where $\sigma = 3 \, (u_c/ u_{0})^3$, $K=(K_y^2 + K_z^2)^{1/2}$, and we consider, for simplicity,
the case when the turbulent Prandtl number ${\rm Pr}_{_{T}}=1$.
Equations~(\ref{AAA3}) and ~(\ref{AAA4}) yield the equation for $\overline{\gamma}= (\gamma + K^2)/\gamma_{0}$ as
\begin{eqnarray}
\overline{\gamma}^3 - \overline{\gamma} - {\pi^2 \, \sigma \, {\rm Sh}^2 \, \alpha^4 \over 2 \epsilon \gamma_0^3 \, (1 + \alpha^2)} =0,
\label{C1}
\end{eqnarray}
where
\begin{eqnarray}
\gamma_{0} &=& {\alpha \over (1+ \alpha^2)^{1/2}} \left[{\rm Ra}_{_{T}} + {\pi^2 \, \sigma  \over 2 \epsilon}\left(1 - \alpha^2\right) \right]^{1/2} ,
\label{C2}
\end{eqnarray}
$\alpha=K_y/K_z$ and $K_z= \pi$.
This implies that $K_y= \alpha \pi$ and $K=(K_y^2 + K_z^2)^{1/2}= \pi \, (1 + \alpha^2)^{1/2}$.
When
\begin{eqnarray}
{\rm Ra}_{_{T}} < {\pi^2 \, \sigma  \over 2 \epsilon}\left(\alpha^2- 1\right)  ,
\label{CCC2}
\end{eqnarray}
$\gamma_{0}$ is a complex function.

\begin{figure}
\vspace*{1mm}
\centering
\includegraphics[width=10.5cm]{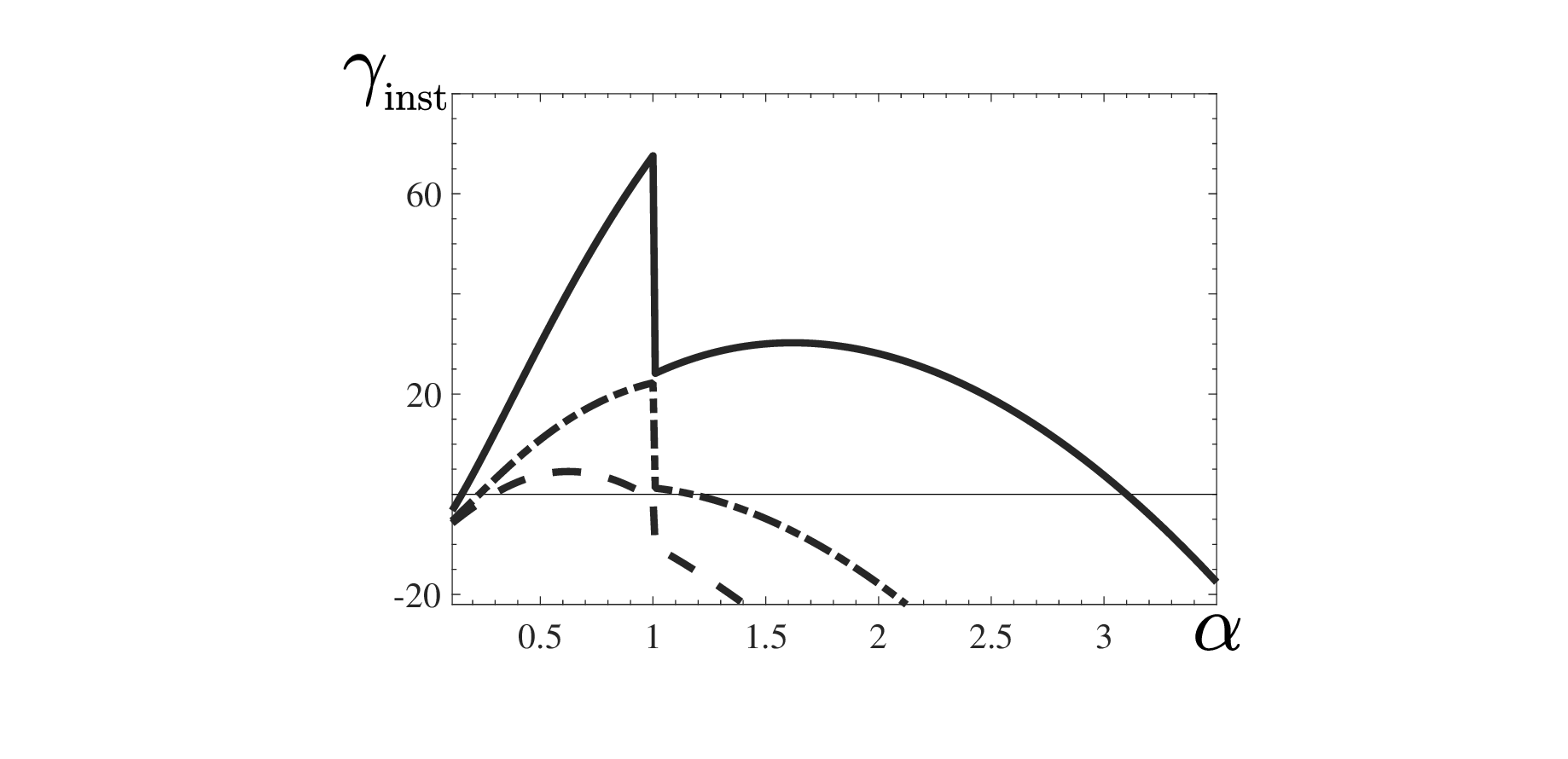}
\includegraphics[width=10.5cm]{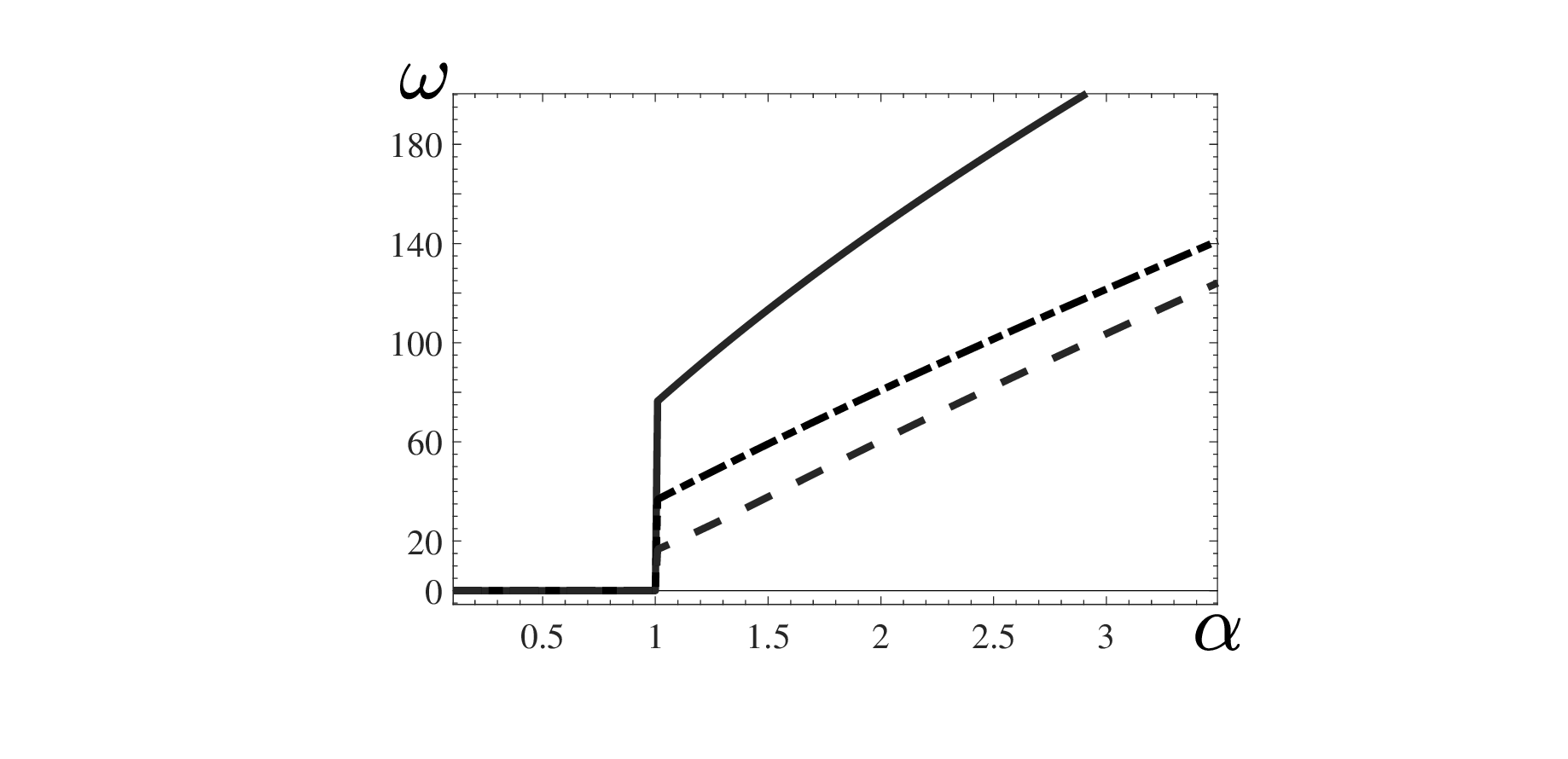}
\caption{\label{Fig4} The growth rate $\gamma_{\rm inst}$ of the instability (upper panel) and the frequency of the excited convective-shear waves $\omega$ (bottom panel) versus the parameter $\alpha$ for the effective Rayleigh number ${\rm Ra}_{_{T}}=0.5$ and different values of the non-dimensional shear number ${\rm Sh}=$ 0.03 (dashed); 0.1 (dashed-dotted); 0.3 (solid).}
\end{figure}

\begin{figure}
\vspace*{1mm}
\centering
\includegraphics[width=10.5cm]{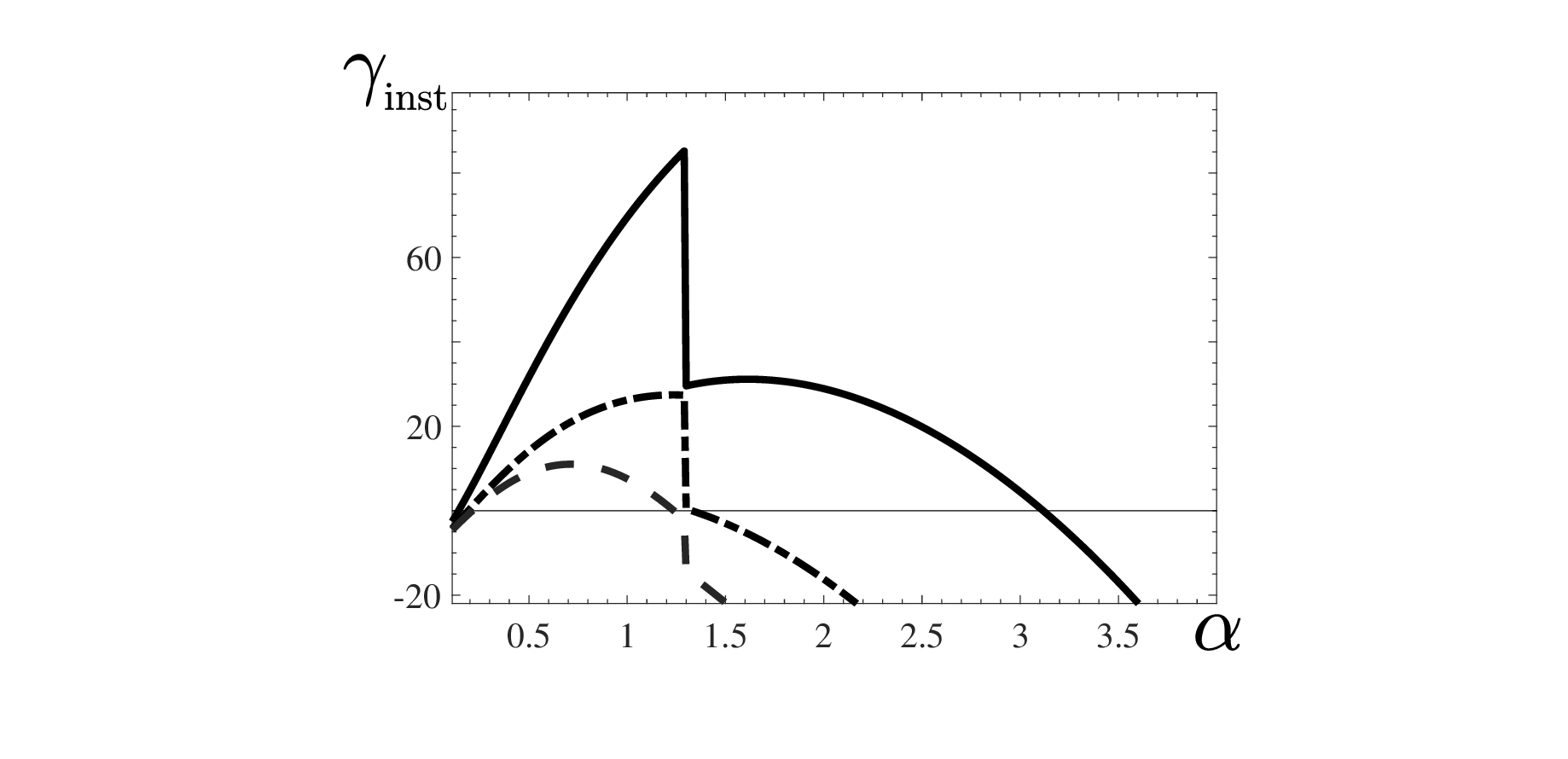}
\includegraphics[width=10.5cm]{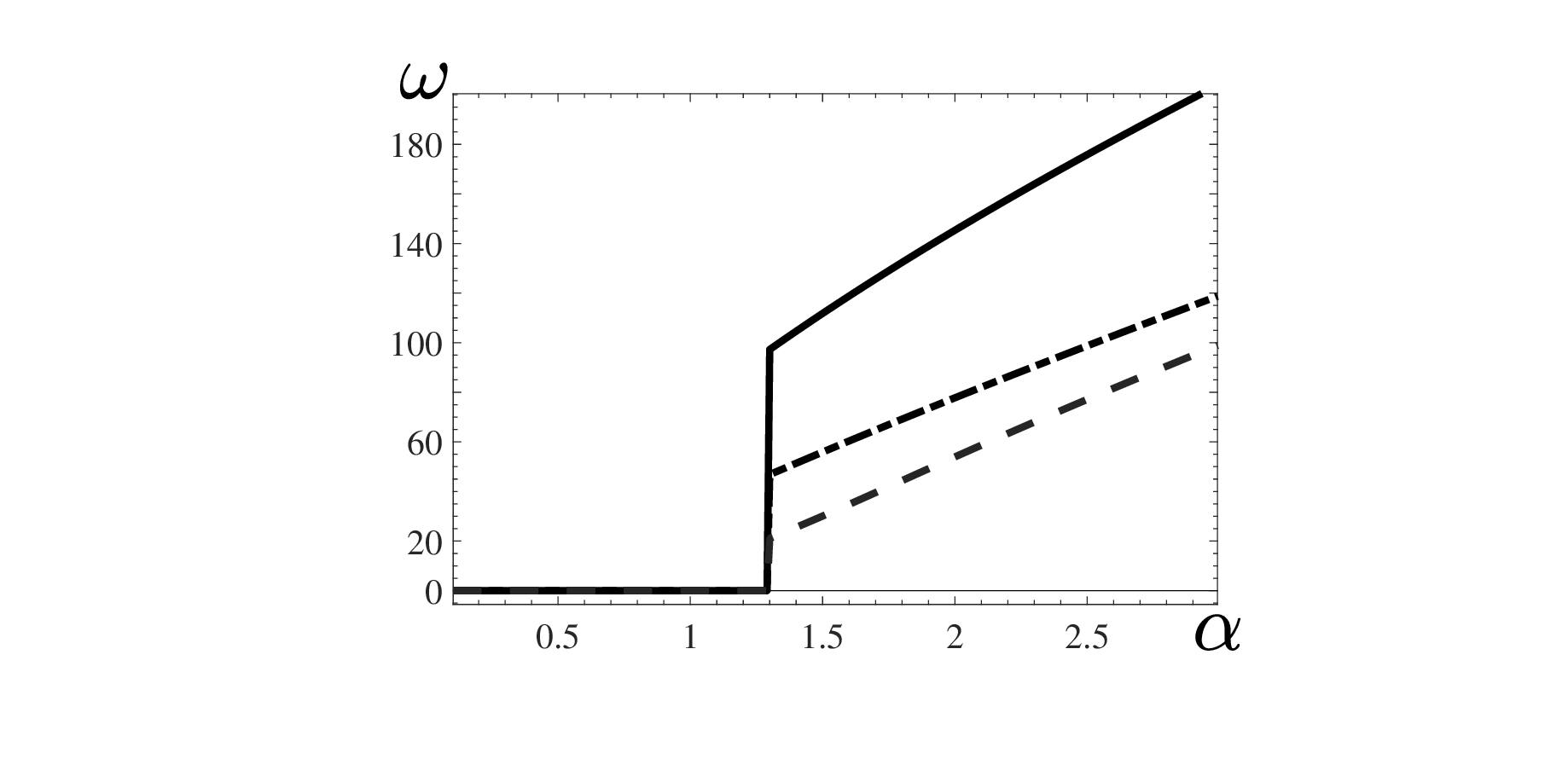}
\caption{\label{Fig5} The growth rate $\gamma_{\rm inst}$ of the instability (upper panel) and the frequency of the excited convective-shear waves $\omega$ (bottom panel) versus the parameter $\alpha$ for the effective Rayleigh number ${\rm Ra}_{_{T}}=10^3$ and different values of the non-dimensional shear number ${\rm Sh}=$ 0.03 (dashed); 0.1 (dashed-dotted); 0.3 (solid).}
\end{figure}

Generally, the solution of the cubic equation~(\ref{C1}) describes two complex conjugate roots
and one real negative root which determine
a damping mode. In the case of the complex conjugate roots, the instability can result in an
excitation of convective-shear waves with the frequency $ \omega = {\rm Im}\{\gamma_{0} \, \overline{\gamma}\}$
and the growth rate  of the instability $ \gamma_{\rm inst} = {\rm Re}\{\gamma_{0} \, \overline{\gamma}\}- \pi^2 (1 + \alpha^2)$, where ${\rm Re}\{Z\}$ is the real part of the complex number and  ${\rm Im}\{Z\}$
is the imagine part of the complex number.

Numerical solution of Eq.~(\ref{C1}) yields the growth rate $\gamma_{\rm inst}$ of the instability and the frequency  $\omega$ of the excited convective-shear waves  versus the parameter $\alpha$ for the effective Rayleigh numbers ${\rm Ra}_{_{T}}=0.5$ (see Fig.~\ref{Fig4}) and ${\rm Ra}_{_{T}}=10^3$ (see Fig.~\ref{Fig5}),
and for different values of the non-dimensional shear number ${\rm Sh}$. It shows that the convective-shear waves are excited for $\alpha>1$ at small effective Rayleigh numbers and for $\alpha>1.3$ at large effective Rayleigh numbers.
The growth rate of the instability and the frequency of the excited convective-shear waves
are very weakly dependent on the effective Rayleigh numbers.
Increase of shear, results in increase of the growth rate $\gamma_{\rm inst}$ of the instability and the frequency of the excited convective-shear waves $\omega$.
The asymptotic solution of Eq.~(\ref{C1}) in the case of $\gamma_{\rm inst} \gg \gamma_{0}$ reads
\begin{eqnarray}
\gamma_{\rm inst} = \left({\pi^2 \, \sigma \, {\rm Sh}^2 \, \alpha^4 \over 2 \epsilon^3 \, (1 + \alpha^2)}\right)^{1/3}  - \pi^2 \, (1 + \alpha^2) ,
\label{C3}
\end{eqnarray}
which corresponds to a non-oscillatory growing solution with $\omega=0$.

\begin{figure}
\vspace*{1mm}
\centering
\includegraphics[width=8cm]{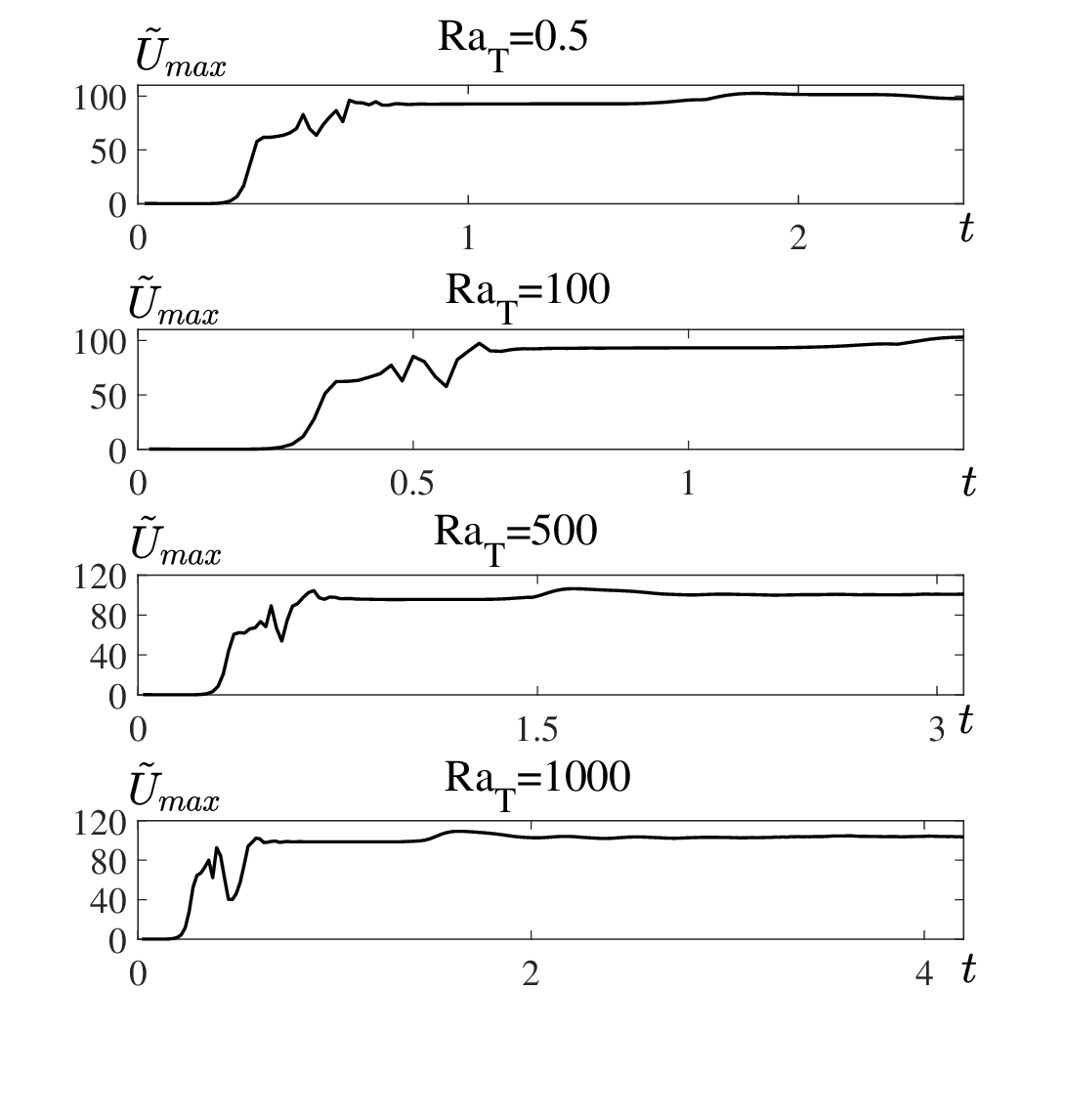}
\caption{\label{Fig6}
Time evolution of the maximum velocity $\tilde {U}_{\rm max}(t)$ for $\epsilon=10^{-3}$, $\alpha=0.5$
and shear number ${\rm Sh}=1$, and at different values of the effective Rayleigh number ${\rm Ra}_{_{T}} =$
0.5; 100; 500 and 1000
for the stress-free boundary conditions.
}
\end{figure}

\begin{figure}
\vspace*{1mm}
\centering
\includegraphics[width=8cm]{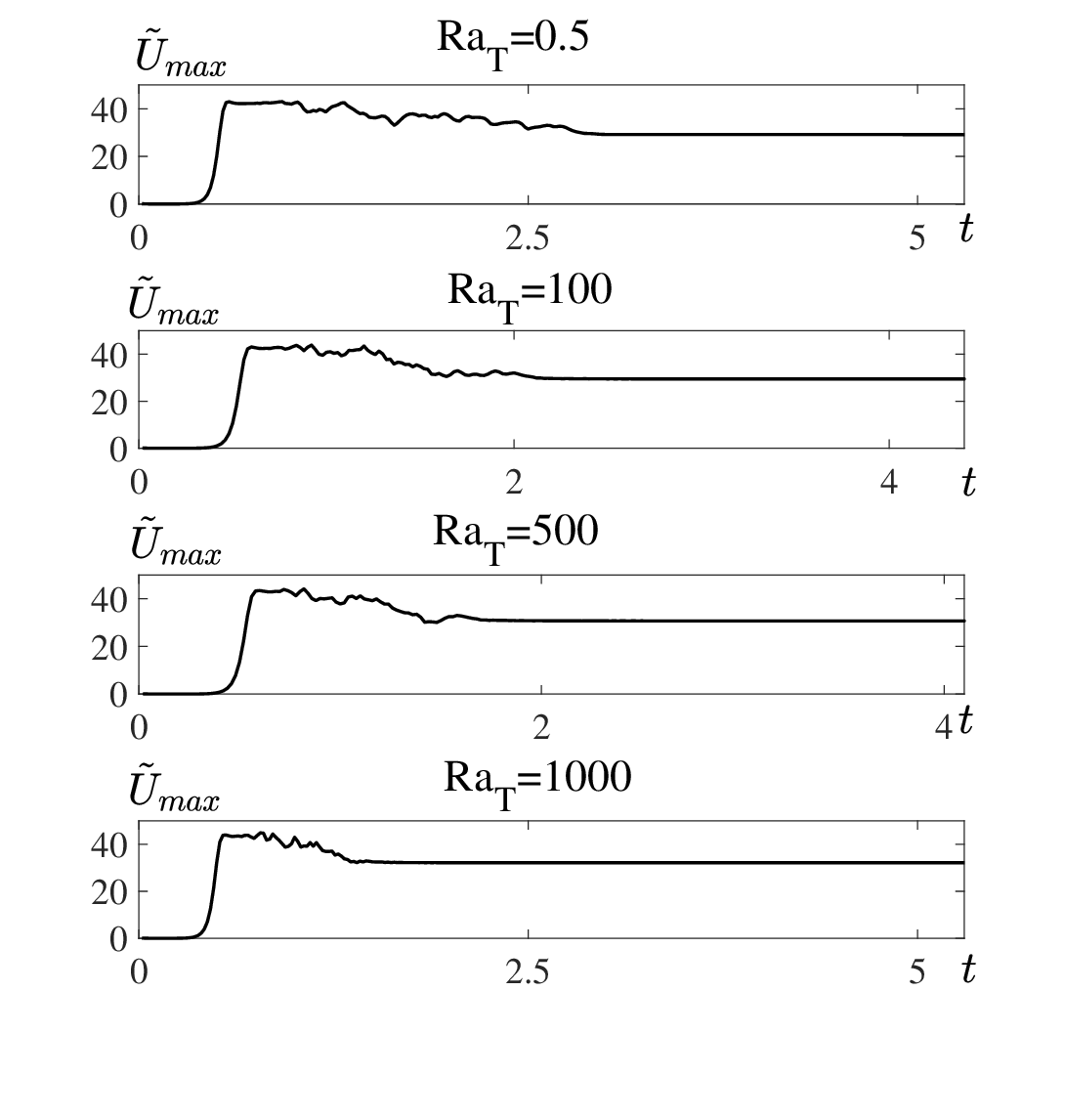}
\caption{\label{Fig7}
Time evolution of the maximum velocity $\tilde {U}_{\rm max}(t)$ for $\epsilon=10^{-3}$, $\alpha=0.5$
and shear number ${\rm Sh}=1$, and at different values of the effective Rayleigh number ${\rm Ra}_{_{T}} =$
0.5; 100; 500 and 1000
for the no-slip boundary conditions.
}
\end{figure}

This instability causes formation of large-scale fluid motions in the form of rolls stretched along the imposed mean wind.
This mechanism can also cause the generation of the convective-shear waves with the frequency shown in
bottom panels of Figs.~\ref{Fig4}--\ref{Fig5}. The convective-shear waves propagate perpendicular to convective rolls.
The predicted motions in convective rolls are characterised by nonzero helicity, in agreement with numerical simulations
(see Ref.~\cite{ET85}).
Note that similar waves propagating in the direction normal to cloud streets
(convective rolls) have been detected in atmospheric convective boundary layers
(see Ref.~\cite{BR99}).
This large-scale instability is fed by the energy of the convective turbulence.

\section{Results of mean-field numerical simulations}

In this section, we discuss results of mean-field numerical simulations
for three-dimensional problem.
We solve numerically Eqs.~(\ref{B1})--(\ref{B2}) for
the periodic boundary conditions in the horizontal $xy$ plane.
The boundary conditions for the potential temperature in the vertical direction are
$\tilde \Theta(t, z=0) = \tilde \Theta(t, z=1)=0$.
We adopt the stress-free and no-slip boundary conditions for the velocity field in the vertical
direction (along the $z$ axis).
The stress-free boundary conditions imply,
\begin{eqnarray}
\tilde U_z(t, z=0) = \tilde U_z(t, z=1)=0 ,
\label{B40}
\end{eqnarray}
\begin{eqnarray}
\nabla_z \tilde U_x(t, z=0) = \nabla_z \tilde U_x(t, z=1)=0 ,
\label{BBB41}
\end{eqnarray}
\begin{eqnarray}
\nabla_z \tilde U_y(t, z=0) = \nabla_z \tilde U_y(t, z=1)=0 ,
\label{B41}
\end{eqnarray}
while the no-slip boundary conditions are given by
\begin{eqnarray}
\tilde {\bm U}(t, z=0) = \tilde {\bm U}(t, z=1)=0 .
\label{BCB43}
\end{eqnarray}

\begin{figure}
\vspace*{1mm}
\centering
\includegraphics[width=8cm]{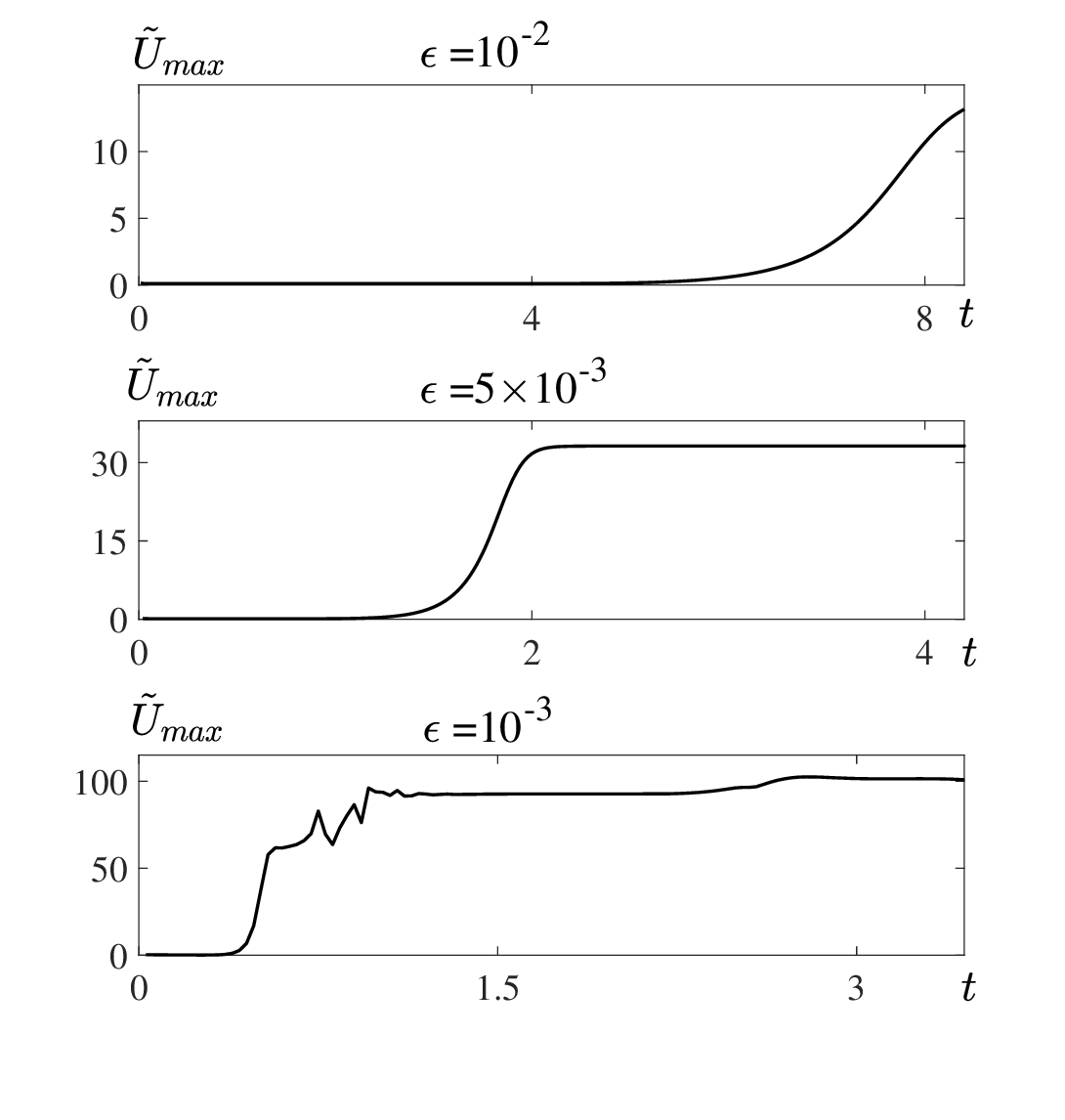}
\caption{\label{Fig8} Time evolution of the maximum velocity $\tilde {U}_{\rm max}(t)$ for
the effective Rayleigh number ${\rm Ra}_{_{T}} =0.5$, $\alpha=0.5$
and shear number ${\rm Sh}=1$, and at different values of the scale separation parameter $\epsilon=10^{-3}; 5 \times 10^{-3}; 10^{-2}$,
for the stress-free boundary conditions.}
\end{figure}

To solve Eqs.~(\ref{B1})--(\ref{B2}), we use the ANSYS FLUENT code (version 19.2)
in the $3D$ box $L_x=5 L_z$ and  $L_y=L_z$,
which is based on the finite volume method.
The discretization method in the code is the second-order.
The simulations are performed with the spatial resolution
$500 \times 100 \times 100$  in $x$, $y$ and $z$ directions, respectively.
A sensitivity check has been also made for the spatial resolution
$500  \times 200 \times 200$.
For both cases similar results for velocity and potential temperature
have been obtained.
In all simulations, we use a time step of $10^{-3}$.
A sensitivity check has been made for time steps as well.
Time steps of $10^{-3}$ and $2 \times 10^{-3}$ have been tested
and a maximum error of 0.05 \%
at velocity and potential temperature between the time steps has been obtained.
In addition, a convergence error was set to be less than $10^{-6}$.

In the mean-field numerical simulations, we use the following values of the
basic dimensionless parameters: the turbulent Prandtl number
${\rm Pr}_{_{T}}=1$, the ratio $u_c / u_{0}=1$, the effective Rayleigh number changes from
${\rm Ra}_{_{T}}=0.5$ to ${\rm Ra}_{_{T}}=1800$ and
the scale separation parameter $\epsilon$ varies from $\epsilon=10^{-3}$ to $\epsilon=10^{-2}$.
The non-dimensional shear number ${\rm Sh}$ varies from ${\rm Sh}=0.01$ to ${\rm Sh}=1$.
Note that the measurements of the two-point correlation function of the turbulent velocity field in the laboratory experiments
with turbulent convection \cite{BEKR09,BEKR11,EKRL23}
and DNS \cite{KKR16,KA24}
show that the integral turbulent scale $\ell_0$ varies from $L_z/10$ to $L_z/5$ which corresponds to
$\epsilon$ used in our mean-field simulations.

\begin{figure}
\vspace*{1mm}
\centering
\includegraphics[width=8cm]{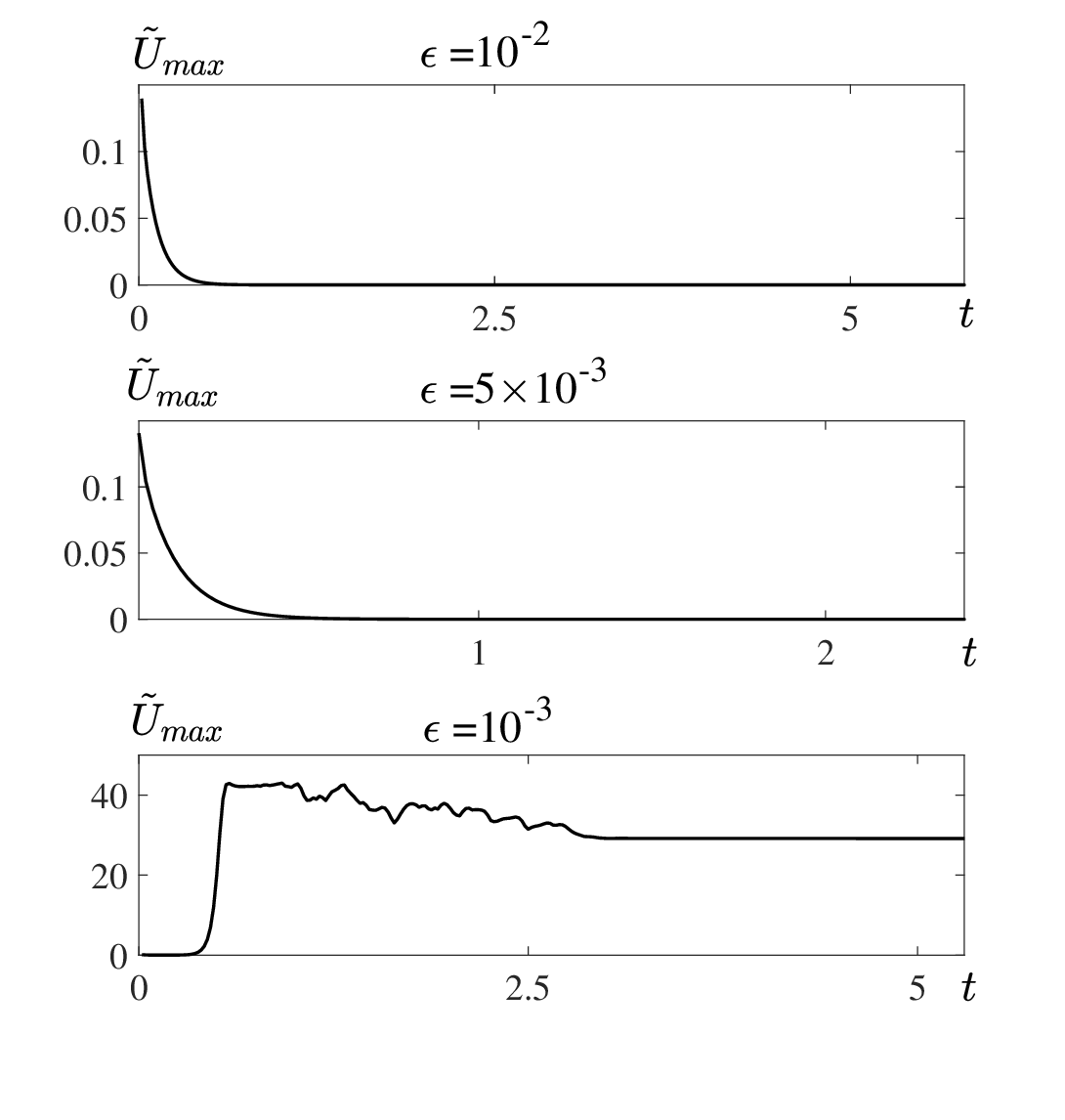}
\caption{\label{Fig9} Time evolution of the maximum velocity $\tilde {U}_{\rm max}(t)$ for
the effective Rayleigh number ${\rm Ra}_{_{T}} =0.5$, $\alpha=0.5$
and shear number ${\rm Sh}=1$, and at different values of the scale separation parameter $\epsilon=10^{-3}; 5 \times 10^{-3}; 10^{-2}$,
for the no-slip boundary conditions.}
\end{figure}

\begin{figure}
\vspace*{1mm}
\centering
\includegraphics[width=8cm]{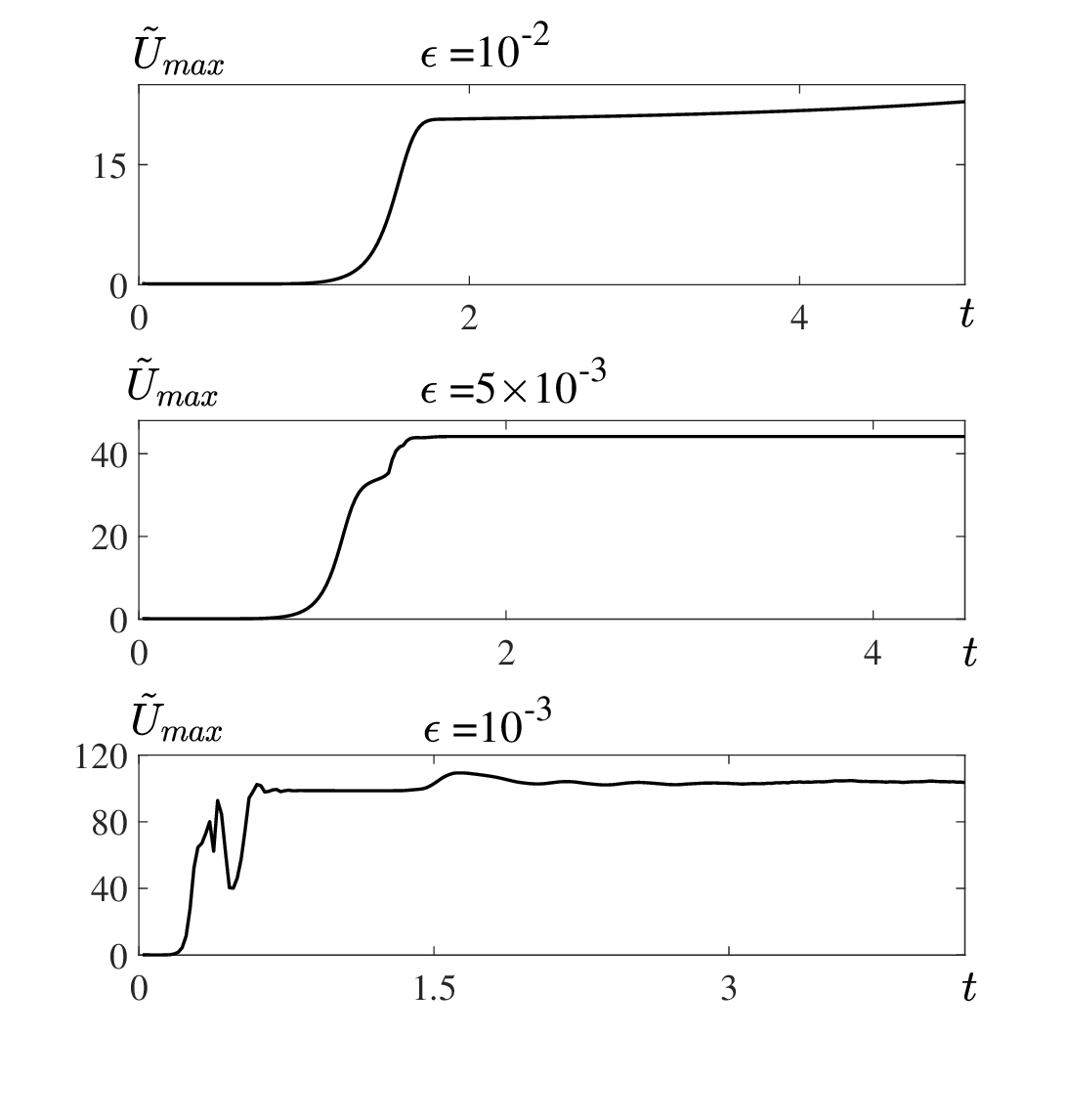}
\caption{\label{Fig10} Time evolution of the maximum velocity $\tilde {U}_{\rm max}(t)$ for
the effective Rayleigh number ${\rm Ra}_{_{T}} =1000$, $\alpha=0.5$
and shear number ${\rm Sh}=1$, and at different values of the scale separation parameter $\epsilon=10^{-3}; 5 \times 10^{-3}; 10^{-2}$,
for the stress-free boundary conditions.}
\end{figure}

\begin{figure}
\vspace*{1mm}
\centering
\includegraphics[width=8cm]{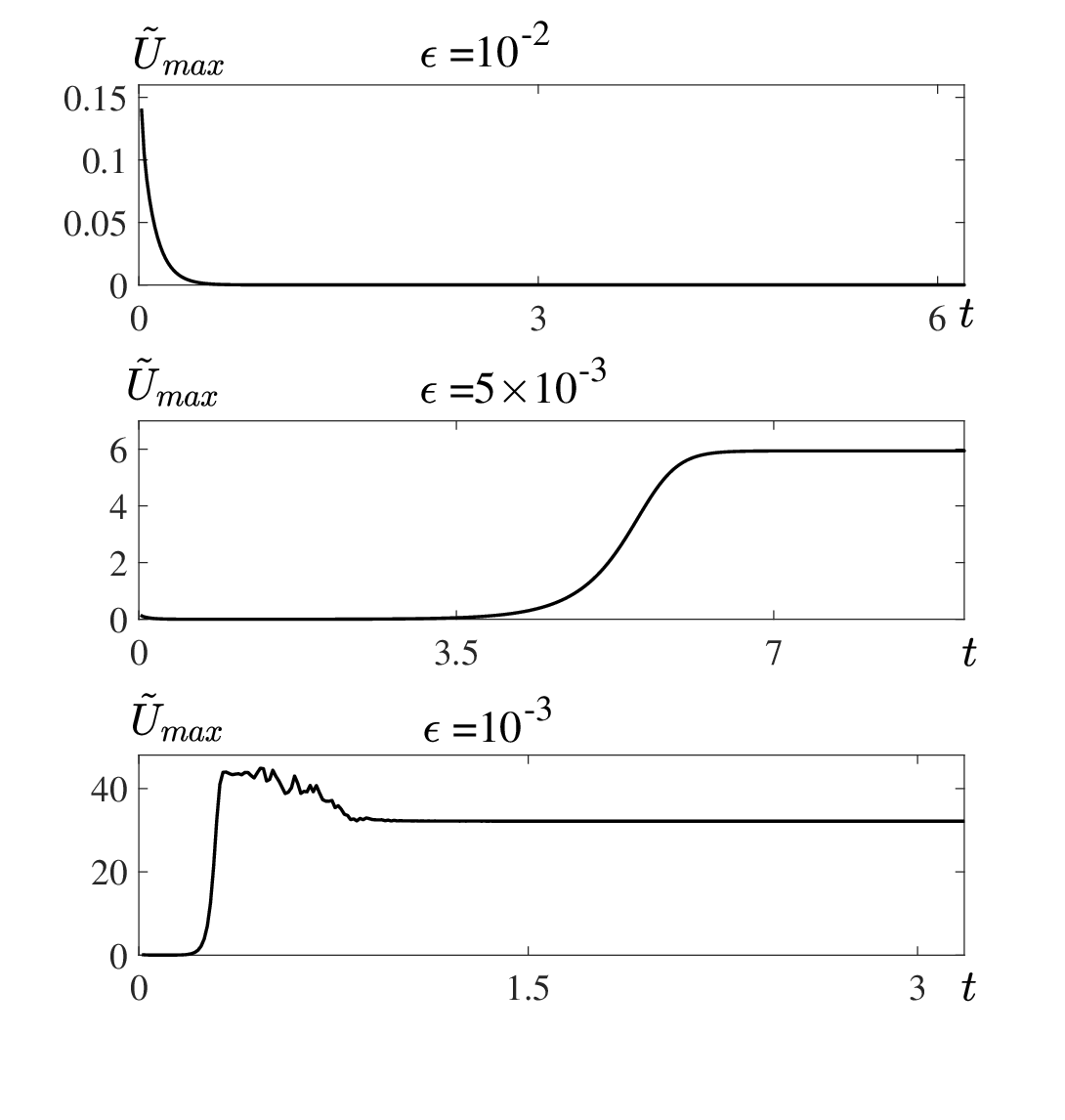}
\caption{\label{Fig11} Time evolution of the maximum velocity $\tilde {U}_{\rm max}(t)$ for
the effective Rayleigh number ${\rm Ra}_{_{T}} =1000$, $\alpha=0.5$
and shear number ${\rm Sh}=1$, and at different values of the scale separation parameter $\epsilon=10^{-3}; 5 \times 10^{-3}; 10^{-2}$,
for the no-slip boundary conditions.}
\end{figure}

For illustration, in Figs.~\ref{Fig6} and~\ref{Fig7} we plot the time evolution of the maximum value $\tilde {U}_{\rm max}(t)$ of the velocity magnitude $|\tilde {\bm U}|$ for different values of the effective Rayleigh numbers ${\rm Ra}_{_{T}}$ changing from  $0.5$
to $10^3$ at a fixed value of the scale separation parameter $\epsilon$
between the vertical size $L_z$ of the computational domain and the integral turbulence scale $\ell_0$
in the sheared large-scale convection (for ${\rm Sh}=1$),
while in Figs.~\ref{Fig8}--\ref{Fig11} we show the time evolution of the maximum velocity $\tilde {U}_{\rm max}(t)$
for different values of the scale separation parameter $\epsilon$ (from  $10^{-3}$ to $10^{-2}$)
(see Figs.~\ref{Fig8}--\ref{Fig9} for ${\rm Ra}_{_{T}}=0.5$
and Figs.~\ref{Fig10}--\ref{Fig11} for ${\rm Ra}_{_{T}}=10^3$).
As can be seen in Figs.~\ref{Fig6}--\ref{Fig11},
at the initial stage of the evolution, the maximum velocity $\tilde {U}_{\rm max}(t)$ increases in time exponentially
due to the excitation of the large-scale convective-shear instability.
The growth rate of the instability predicted from the theory [as the solution of the cubic
dispersion equation~(\ref{C1})] is in agreement with that obtained in the mean-field numerical simulations.

During the nonlinear stage of the instability, we observe that
$\tilde {U}_{\rm max}(t)$ reaches the maximum value
which is  weakly dependent on the  effective Rayleigh number ${\rm Ra}_{_{T}}$
for the stress-free and no-slip boundary conditions
(see Figs.~\ref{Fig6}--\ref{Fig7}).
On the other hand, the maximum value of the function $\tilde {U}_{\rm max}$ at the stationary stage
strongly depends on the scale separation parameter $\epsilon$
and on the boundary conditions (see Figs.~\ref{Fig8}--\ref{Fig9}).
In particular, increasing the scale separation between the vertical size $L_z$ of the computational domain
and the integral turbulence scale $\ell_0$ (i.e., decreasing the parameter $\epsilon$),
the maximum value of the function $\tilde {U}_{\rm max}$ increases.
For the no-slip vertical boundary conditions, the large-scale instability is excited
and convective structures are formed
when $\epsilon < 3 \times 10 ^{-3}$ at ${\rm Ra}_{_{T}} = 0.5$,
$\epsilon < 4 \times 10 ^{-3}$ at ${\rm Ra}_{_{T}} = 500$
and $\epsilon < 7 \times 10 ^{-3}$ at ${\rm Ra}_{_{T}} = 10^3$.
In addition, the time evolution in the nonlinear stage of the instability for
the stress-free and no-slip boundary conditions are different.

Some features in the time evolution of the maximum velocity $\tilde {U}_{\rm max}(t)$
have been also observed in the recent mean-field simulations of the shear-free convection
(see Ref.~\cite{OKR22}), where $\tilde {U}_{\rm max}(t)$ is independent of ${\rm Ra}_{_{T}}$
for the same boundary conditions, but it strongly depends on the scale separation parameter $\epsilon$.
However, in a sheared large-scale convection, we do not observe clear
nonlinear oscillations of $\tilde {U}_{\rm max}$ which have been seen after the steady-state stage
(when the function $\tilde {U}_{\rm max}$ is nearly constant in time)
in the large-scale shear-free convection (see Ref.~\cite{OKR22}).

\begin{figure}
\vspace*{1mm}
\centering
\includegraphics[width=9cm]{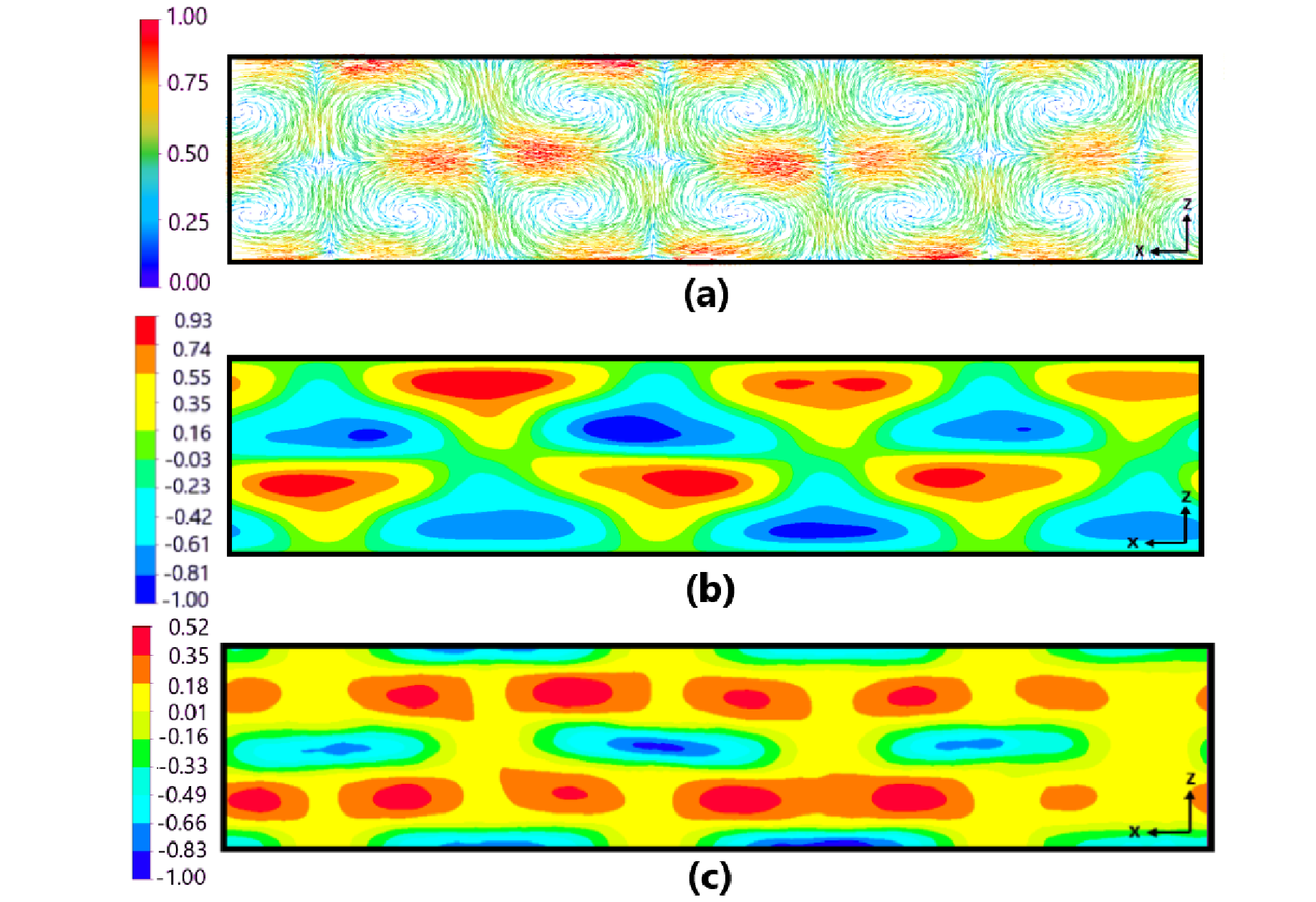}
\caption{\label{Fig12} The patterns of velocity vectors $\tilde {\bm U}=\tilde {U}_x \, {\bm e}_x + \tilde {U}_z \, {\bm e}_z$
with $\tilde {U}_y \to 0$ (Fig.~\ref{Fig12}a);
the patterns of the potential temperature deviations $\tilde \Theta \, {\rm Ra}_{_{T}}$
from the equilibrium potential temperature in the basic reference state (Fig.~\ref{Fig12}b);
the patterns of the vertical gradient of the mean potential temperature $(\nabla_z \tilde \Theta -1)
\, {\rm Ra}_{_{T}}$ (Fig.~\ref{Fig12}c) at time instant $t=0.34$ of the turbulent viscosity time $L_z^2/\nu_{_{T}}$,
effective Rayleigh number ${\rm Ra}_{_{T}}=0.5$, $\epsilon=10^{-3}$, $\alpha=0.5$
and shear number ${\rm Sh}=1$ for the stress-free vertical boundary conditions.
All quantities are normalized by their maximum values.}
\end{figure}

\begin{figure}
\vspace*{1mm}
\centering
\includegraphics[width=7cm]{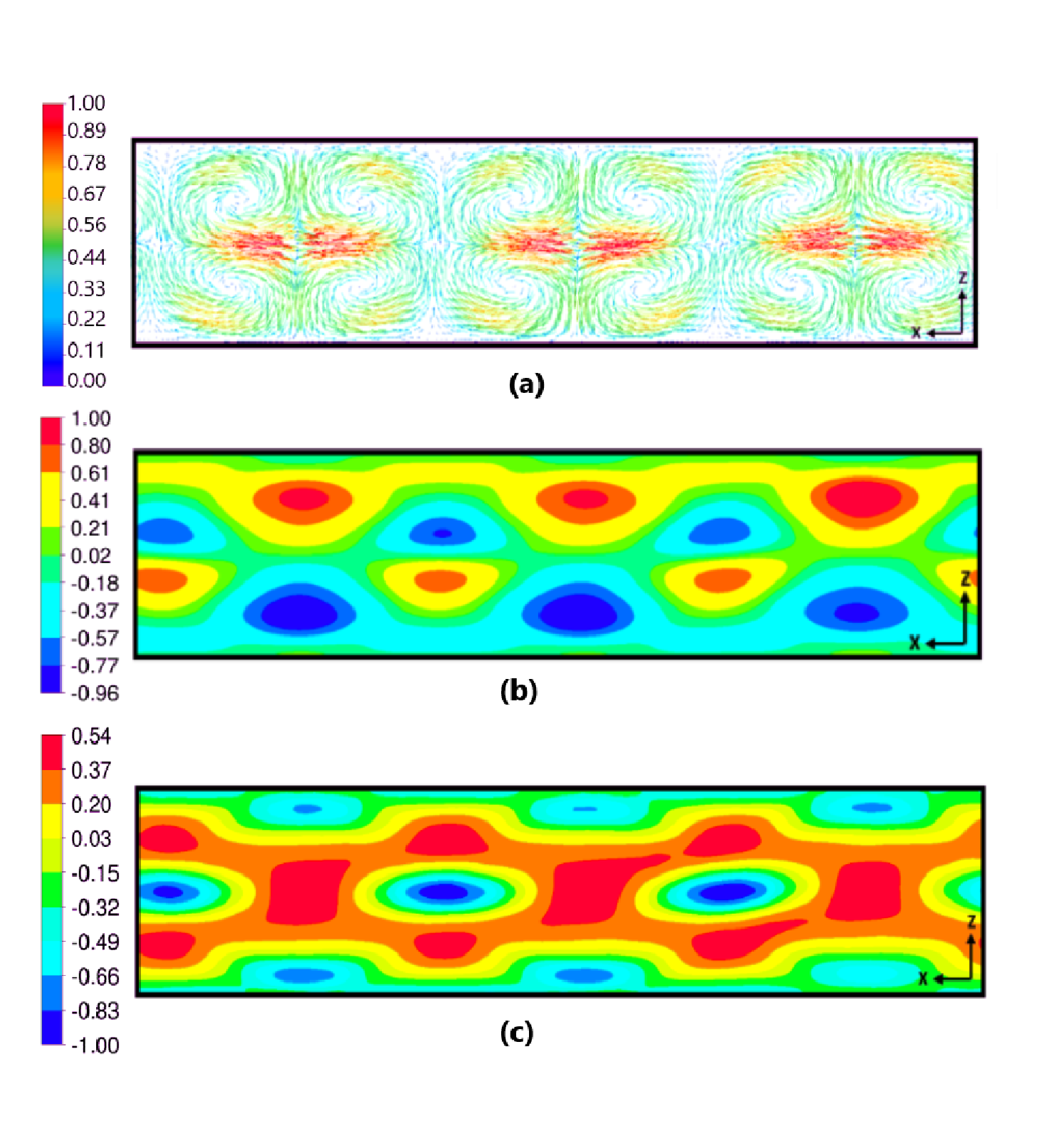}
\caption{\label{Fig13} The patterns of velocity vectors $\tilde {\bm U}=\tilde {U}_x \, {\bm e}_x + \tilde {U}_z \, {\bm e}_z$
with $\tilde {U}_y \to 0$  (Fig.~\ref{Fig13}a);
the patterns of the potential temperature deviations $\tilde \Theta \, {\rm Ra}_{_{T}}$
from the equilibrium potential temperature in the basic reference state (Fig.~\ref{Fig13}b);
the patterns of the vertical gradient of the mean potential temperature
$(\nabla_z \tilde \Theta -1) \, {\rm Ra}_{_{T}}$ (Fig.~\ref{Fig13}c) at time instant $t=0.7$
of the turbulent viscosity time $L_z^2/\nu_{_{T}}$,
effective Rayleigh number ${\rm Ra}_{_{T}}=0.5$, $\epsilon=10^{-3}$, $\alpha=0.5$
and shear number ${\rm Sh}=1$ for the no-slip vertical boundary conditions.
All quantities are normalized by their maximum values.}
\end{figure}

\begin{figure}
\vspace*{1mm}
\centering
\includegraphics[width=9cm]{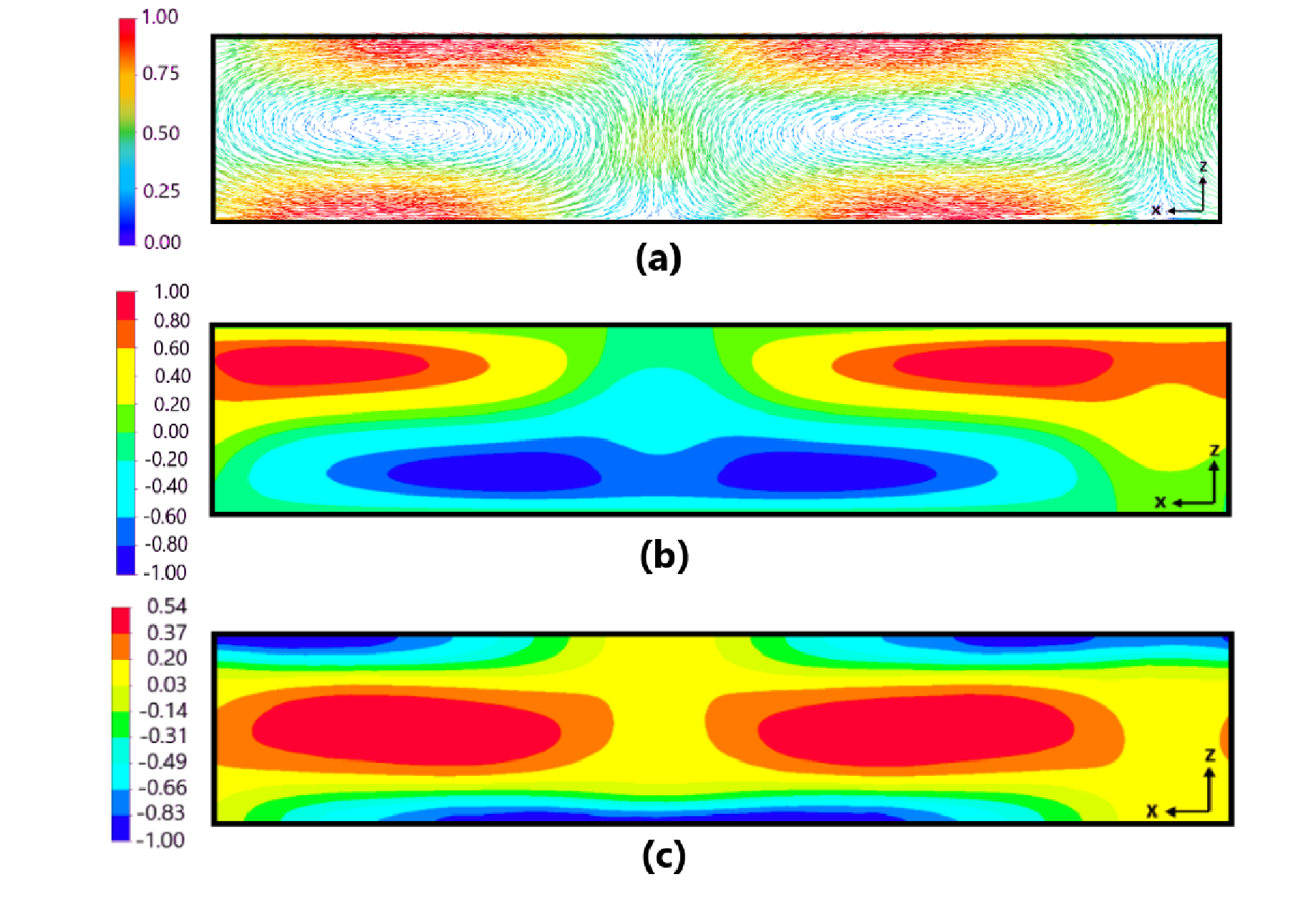}
\caption{\label{Fig14} The patterns of velocity vectors $\tilde {\bm U}=\tilde {U}_x \, {\bm e}_x + \tilde {U}_z \, {\bm e}_z$
with $\tilde {U}_y \to 0$  (Fig.~\ref{Fig14}a);
the patterns of the potential temperature deviations $\tilde \Theta \, {\rm Ra}_{_{T}}$ from the equilibrium potential temperature in the basic reference state (Fig.~\ref{Fig14}b);
the patterns of the vertical gradient of the mean potential temperature $(\nabla_z \tilde \Theta -1) \, {\rm Ra}_{_{T}}$ (Fig.~\ref{Fig14}c) at time instant $t=1$ of the turbulent viscosity time $L_z^2/\nu_{_{T}}$,
effective Rayleigh number ${\rm Ra}_{_{T}}=0.5$, $\epsilon=10^{-3}$, $\alpha=0.5$
and shear number ${\rm Sh}=1$ for the stress-free vertical boundary conditions.
All quantities are normalized by their maximum values.
}
\end{figure}

\begin{figure}
\vspace*{1mm}
\centering
\includegraphics[width=7cm]{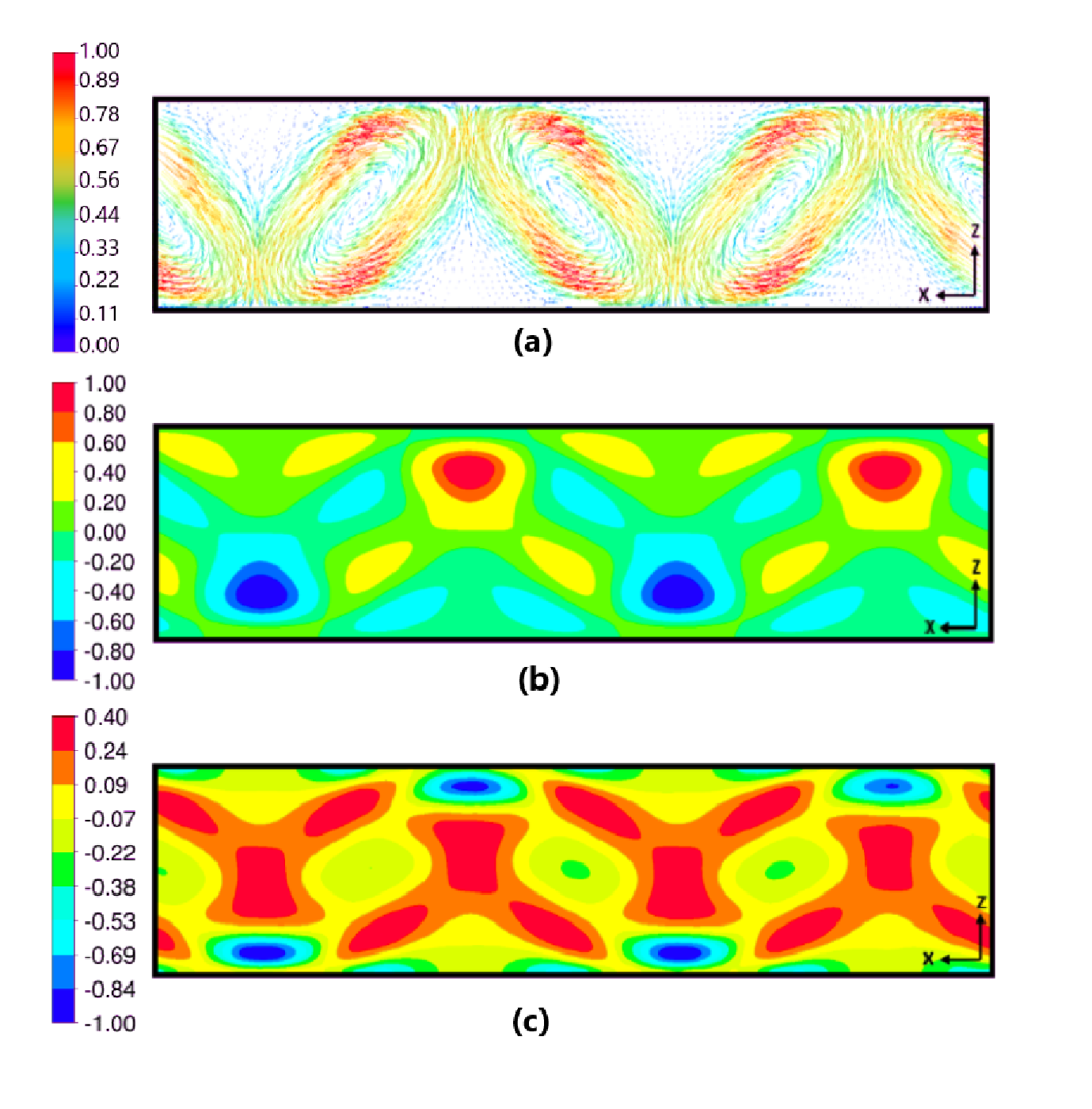}
\caption{\label{Fig15} The patterns of velocity vectors $\tilde {\bm U}=\tilde {U}_x \, {\bm e}_x + \tilde {U}_z \, {\bm e}_z$
with $\tilde {U}_y \to 0$  (Fig.~\ref{Fig15}a);
the patterns of the potential temperature deviations $\tilde \Theta \, {\rm Ra}_{_{T}}$ from the equilibrium potential temperature in the basic reference state (Fig.~\ref{Fig15}b);
the patterns of the vertical gradient of the mean potential temperature $(\nabla_z \tilde \Theta -1) \, {\rm Ra}_{_{T}}$ (Fig.~\ref{Fig15}c) at time instant $t=4$ of the turbulent viscosity time $L_z^2/\nu_{_{T}}$,
effective Rayleigh number ${\rm Ra}_{_{T}}=0.5$, $\epsilon=10^{-3}$, $\alpha=0.5$
and shear number ${\rm Sh}=1$ for the no-slip vertical boundary conditions.
All quantities are normalized by their maximum values.}
\end{figure}

\begin{figure}
\vspace*{1mm}
\centering
\includegraphics[width=9cm]{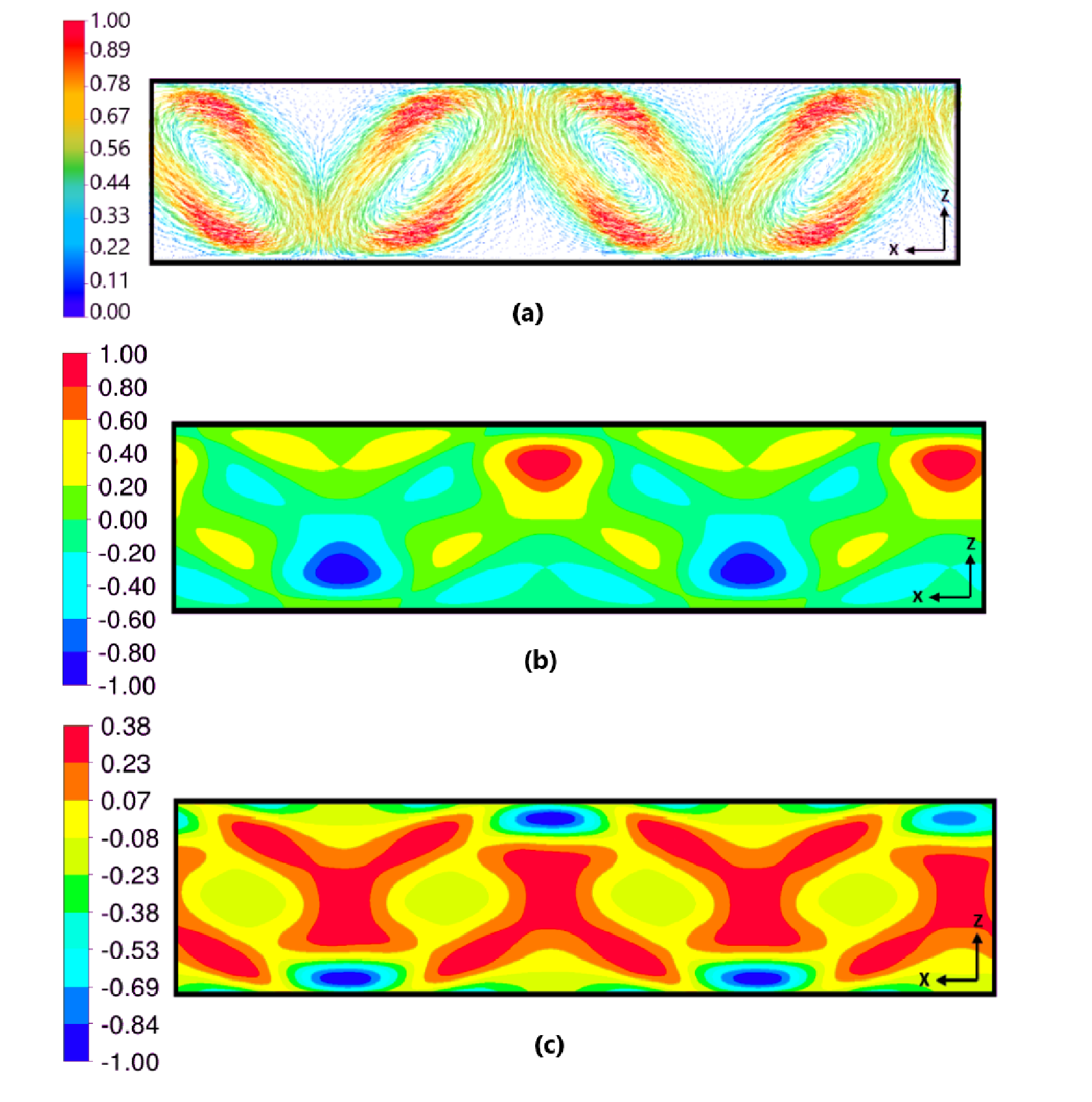}
\caption{\label{Fig16} The patterns of velocity vectors $\tilde {\bm U}=\tilde {U}_x \, {\bm e}_x + \tilde {U}_z \, {\bm e}_z$
with $\tilde {U}_y \to 0$  (Fig.~\ref{Fig16}a);
the patterns of the potential temperature deviations $\tilde \Theta \, {\rm Ra}_{_{T}}$ from the equilibrium potential temperature in the basic reference state  (Fig.~\ref{Fig16}b);
the patterns of the vertical gradient of the mean potential temperature $(\nabla_z \tilde \Theta -1) \, {\rm Ra}_{_{T}}$  (Fig.~\ref{Fig16}c) at time instant $t=5.32$ of the turbulent viscosity time $L_z^2/\nu_{_{T}}$,
effective Rayleigh number ${\rm Ra}_{_{T}}=1000$, $\epsilon=10^{-3}$, $\alpha=0.5$
and shear number ${\rm Sh}=1$ for the no-slip vertical boundary conditions.
All quantities are normalized by their maximum values.
}
\end{figure}

To observe spatial structure of the basic characteristics of the sheared large-scale
convection, in Figs.~\ref{Fig12}--\ref{Fig16}
we plot the patterns of velocity vectors $\tilde {\bm U}=\tilde {U}_y \, {\bm e}_y + \tilde {U}_z \, {\bm e}_z$
with $\tilde {U}_x \to 0$ (panels a);
the patterns of the potential temperature deviations $\tilde \Theta \, {\rm Ra}_{_{T}}$
from the equilibrium potential temperature in the basic reference state (panels b) and
the patterns of the vertical gradient of the mean potential temperature $(\nabla_z \tilde \Theta -1) \, {\rm Ra}_{_{T}}$
(panels c) at several time instants:
before the system reaches steady-state (see Figs.~\ref{Fig12} and~\ref{Fig13})
and after the system reaches steady-state  (see Fig.~\ref{Fig14}--\ref{Fig16}).
Here ${\bm e}_x$,  ${\bm e}_y$ and ${\bm e}_z$ are the unit vectors directed along the $x$, $y$ and $z$ axes.

We remind that the potential temperature is measured
in the units of $L_z \, N^2 \, {\rm Pr}_{_{T}}/\beta
= {\rm Ra}_{_{T}} \nu_{_{T}}^2/(\beta L_z^3)$. This is the
reason why we show in Figs.~\ref{Fig12}--\ref{Fig16} (see panels b)
the pattern of the normalized deviations of the potential
temperature $\tilde \Theta \, {\rm Ra}_{_{T}}$ from the equilibrium potential
temperature in the basic reference state. Note also that the total gradient
of the potential temperature is the sum of the equilibrium
constant gradient of the potential temperature
$\nabla_z \meanT_{\rm eq}$ (negative for a convection) and the gradient of
the potential temperature $\nabla_z \meanTheta$. Therefore, we show in
Figs.~\ref{Fig12}--\ref{Fig16} (see panels c) the pattern of the normalized
total vertical gradient of the mean potential temperature,
$(\nabla_z \tilde \Theta -1) \, {\rm Ra}_{_{T}}$ which characterises the large-scale
convection.

\begin{figure}
\vspace*{1mm}
\centering
\includegraphics[width=6cm]{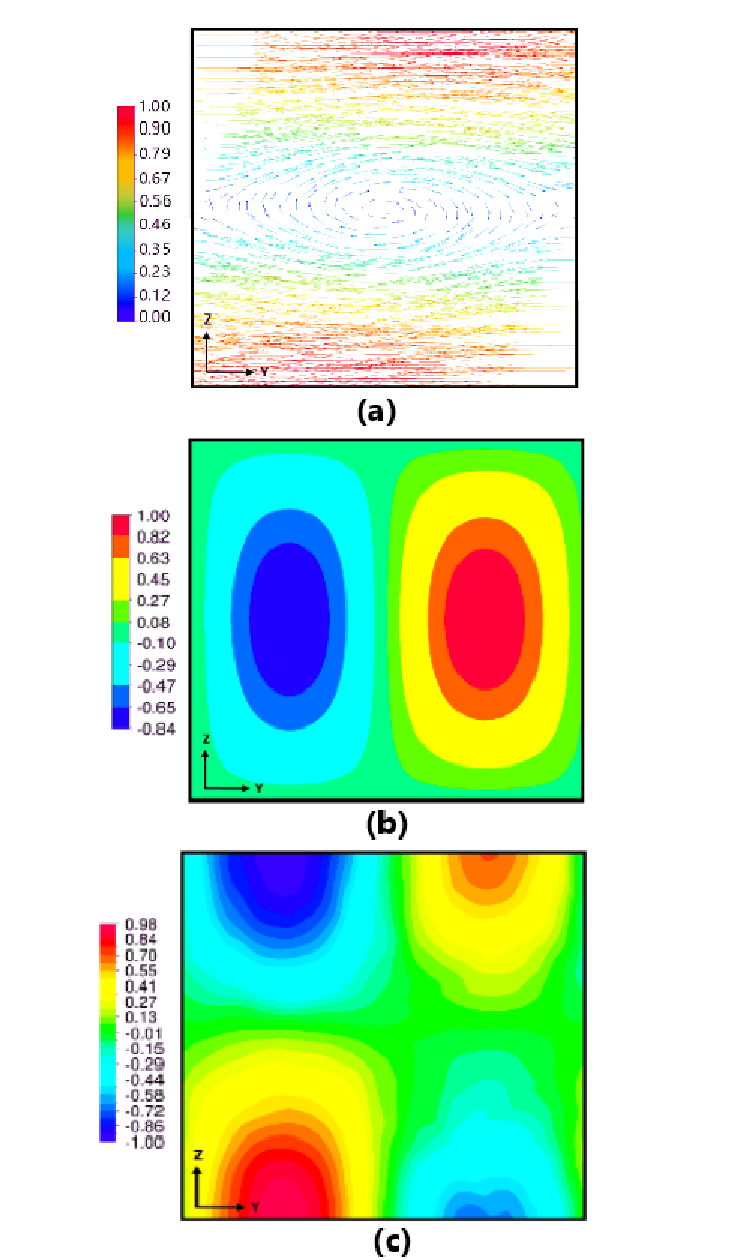}
\caption{\label{Fig17} The patterns of velocity vectors $\tilde {\bm U}=\tilde {U}_y \, {\bm e}_y + \tilde {U}_z \, {\bm e}_z$
with $\tilde {U}_x \to 0$   (Fig.~\ref{Fig17}a);
the patterns of the potential temperature deviations $\tilde \Theta \, {\rm Ra}_{_{T}}$
from the equilibrium potential temperature in the basic reference state  (Fig.~\ref{Fig17}b);
the patterns of the vertical gradient of the mean potential temperature $(\nabla_z \tilde \Theta -1) \, {\rm Ra}_{_{T}}$
(Fig.~\ref{Fig17}c) at time instant $t=0.04$ of the turbulent viscosity time $L_z^2/\nu_{_{T}}$,
effective Rayleigh number ${\rm Ra}_{_{T}}=0.5$, $\epsilon=10^{-3}$, $\alpha=0.5$
and shear number ${\rm Sh}=1$ for the stress-free vertical boundary conditions.
All quantities are normalized by their maximum values.}
\end{figure}

At the linear stage of the system evolution, the patterns for the stress-free and no-slip boundary conditions
are the same (see Fig.~\ref{Fig17} shown in the $yz$ plane, where in the panel (a) we plot the patterns of velocity vectors $\tilde {\bm U}=\tilde {U}_y \, {\bm e}_y + \tilde {U}_z \, {\bm e}_z$
with $\tilde {U}_x \to 0$).
During the evolution, there is a transition from the large-scale circulations
at the $yz$ plane seen in the linear stage of the instability  (see Fig.~\ref{Fig17})
to the the large-scale circulations  at the $xz$ plane observed during nonlinear stage of the instability (see Figs.~\ref{Fig12}--\ref{Fig13}).
In addition, there is a transition from the two-layer vertical structure of the mean velocity field
with two convective rolls in the $z$ direction and six rolls in the $x$ direction
(as can be seen in Fig.~\ref{Fig12}  for the stress-free boundary conditions)
to  the one-layer vertical structure with one roll in the $z$ direction
and two rolls in the $x$ direction (see Fig.~\ref{Fig14}).

On the other hand, comparing Figs.~\ref{Fig13} and~\ref{Fig15} (which corresponds to the no-slip boundary conditions),
we observe that during nonlinear stage of the instability there is a transition from the large-scale circulations
with the two-layer vertical structure with two convective rolls in the $z$ direction
and six rolls in the $x$ direction (see Fig.~\ref{Fig13}) to
the one-layer structure with four inclined rolls (see Fig.~\ref{Fig15}).
The formation of the inclined rolls for the no-slip boundary conditions
are already observed starting with $t = 0.9$ of the turbulent viscosity time $L_z^2/\nu_{_{T}}$
at  the  effective Rayleigh number ${\rm Ra}_{_{T}}=0.5$.
The inclined rolls for the no-slip boundary conditions are also observed at  the  effective Rayleigh number ${\rm Ra}_{_{T}}=10^3$
(see Fig.~\ref{Fig16}).

As has been observed in the shear-free convection \cite{OKR22}, the large-scale convective structures are also formed
in a sheared convection even at low values of the effective Rayleigh numbers ${\rm Ra}_{_{T}} = 0.5$
due to the additional terms proportional to $\epsilon$ in Eq.~(\ref{B2}) for the evolution of the potential
temperature (which are caused by the modification of the turbulent heat flux by non-uniform fluid flows).
Moreover, in Figs.~\ref{Fig12}--\ref{Fig16} (panels c), one can see the regions with the positive gradient of the potential temperature $(\nabla_z \tilde \Theta -1) \, {\rm Ra}_{_{T}}$, which are typical for stably stratified turbulence.
Such effects have been previously observed in experiments \cite{NSS01,BEKR20}, direct numerical simulations \cite{AB16,KRB17,TSW17,KA19} of turbulent convection and shear-free mean-field numerical simulations \cite{OKR22}.
The formation of the regions with the positive gradient of the potential temperature
inside the large-scale circulation can be understood as follows.
The total vertical heat flux ${\bm F}_z^{\rm tot}$ includes three contributions \cite{OKR22}:
\begin{itemize}
\item{
the mean vertical heat flux $\meanUU_z \, \meanTheta$ of the
large-scale circulation,}
\item{
the vertical turbulent heat flux
${\bm F}_{z}^\ast=- \kappa_{_{T}} {\bm \nabla}_z \meanTheta$, and}
\item{
the new turbulent heat flux ${\bm F}_z^{\rm new} = - \tau_0 \, {\bm F}_{z}^{\ast} \, {\rm div}
\, \meanUU_{\perp}$.}
\end{itemize}
Therefore, the vertical gradient $\nabla_z \meanTheta$ of the mean potential temperature is given by
\begin{eqnarray}
\nabla_z \meanTheta = {\meanU_z \, \meanTheta - F_z^{\rm tot} \over \kappa_{_{T}} \,
(1 - \tau_0 \, {\rm div}\, \meanUU_{\perp})}  .
\label{F1}
\end{eqnarray}
Inside the large-scale circulation where $\meanU_z \, \meanTheta > F_z^{\rm tot}$,
the vertical gradient $\nabla_z \meanTheta$ is positive.
On the other hand, when $\meanU_z \, \meanTheta < F_z^{\rm tot}$,
the vertical gradient $\nabla_z \meanTheta$ is negative.
Here we take into account that $\tau_0 \, |{\rm div}\, \meanUU_{\perp}| < 1$.

In the present study, we also observe the formation the large-scale rolls even below the threshold of the laminar convection
(see Figs.~\ref{Fig12}--\ref{Fig15} for effective Rayleigh number ${\rm Ra}_{_{T}}=0.5$
which can be compared  with Fig.~\ref{Fig16} for ${\rm Ra}_{_{T}}=1000$).
This is because turbulence with non-uniform large-scale flows contributes to the
turbulent heat flux.
The reasons for this effect are related to production of additional essentially anisotropic velocity fluctuations generated by tangling of the mean-velocity gradients by small-scale turbulent motions due to the influence of the inertial forces during the lifetime of turbulent eddies.
These anisotropic velocity fluctuations contribute to the turbulent heat flux.
In particular, anisotropic velocity fluctuations ${\bm u}^{\rm anisot}$ are produced
by the large-scale shear $\nabla_j \meanU_{i}$
in the presence of isotropic velocity fluctuations ${\bm u}^{\rm isot}$, i.e.,
$\partial  {\bm u}^{\rm anisot} / \partial t \propto - ({\bm u}^{\rm isot} \cdot \bec{\nabla}) \meanU_{i}$.
This mechanism is called "tangling".
Similar mechanism exists in fluid dynamics, when temperature fluctuations $\theta$ are produced by
"tangling" of the gradient of the mean temperature $\nabla_j \meanT$ by velocity fluctuations ${\bm u}$, i.e.,
$\partial  \theta / \partial t \propto -  ({\bm u} \cdot \bec{\nabla}) \meanT$.
Moreover, the tangling mechanism also exists in magnetohydrodynamics,
when magnetic fluctuations ${\bm b}$ are produced by
"tangling" of the mean magnetic field $\meanBB$ by velocity fluctuations, i.e.,
$\partial  {\bm b} / \partial t \propto (\meanBB \cdot \bec{\nabla}) {\bm u}$.
So the "tangling" is an universal mechanism of production of anisotropic fluctuations.

The modification of the turbulent heat flux by anisotropic
velocity fluctuations causes an excitation of large-scale convective-shear instability,
which results in the formation of large-scale semi-organized structures in the form of
rolls and generation of convective-shear waves propagating perpendicular to the convective rolls.
The life-times and spatial scales of these structures are much larger compared to the largest turbulent time scales.
As the result, the evolutionary equation~(\ref{B2}) for the potential temperature $\tilde \Theta$
contains the new terms proportional to the spatial derivatives of the mean velocity field $\tilde {\bm U}$ (see the terms proportional to the parameter $\epsilon$).

The developed mean-field theory which yields a convective-shear instability is
capable to explain the properties of semi-organized structures observed in the atmospheric convective
boundary layer.
In particular, in the presence of the mean wind,
the semi-organized structures are observed in the
form of rolls (cloud streets). The observed rolls usually align along the mean horizontal wind of the convective
layer, with lengths varying from $20$ to $200$ km, widths changing from $2$ to $10$
km, and convective depths varying from $2$ to $3$ km \cite{AZ96}.
The typical value of the aspect ratio of the observed rolls is $L^{\rm cs}_{z} / L^{\rm cs}_{\rm h} \approx
0.14$ -- $1$, where $L^{\rm cs}_{z}$ and $L^{\rm cs}_{\rm h}$ are the vertical and horizontal scales of the cloud streets
in the plane perpendicular to the mean wind.
The ratio of the minimal size $L^{\rm cs}$ of the observed structures to the integral
scale of turbulent motions is $L^{\rm cs} / \ell_{0} = 10$ -- $100$. The
characteristic life time of rolls varies from one to tens hours.
The suggested theory predicts the following parameters
of the convective rolls: the aspect ratio $L^{\rm cs}_{z} / L^{\rm cs}_{\rm h}$
ranges from very small values to  1, and $ L^{\rm cs} / \ell_{0}$ varies from 10 to100.
The characteristic time of formation of the rolls $\sim \tau_0 /
\gamma_{\rm inst} $ varies from 1 to 3 hours.
It is of the order of several turbulent viscosity time.

\section{Conclusions}

Mean-field simulations based on a developed mean-field theory \cite{EKRZ02,EKRZ06}
of a non-rotating sheared turbulent convection are performed.
This mean-field theory describes an effect of modification of the turbulent heat flux
by the non-uniform large-scale motions caused by production of essentially anisotropic
velocity fluctuations generated by tangling of the mean-velocity gradients by small-scale turbulent motions.
The effect causes an excitation of large-scale convective-shear instability and the formation of large-scale convective rolls.
During the nonlinear stage of the convective-shear instability,
there is a transition from the two-layer vertical structure with two rolls in the vertical direction
before the system reaches steady-state to  the one-layer vertical structure
with one roll after the system reaches steady-state.
This effect is observed for all effective Rayleigh numbers.

The performed mean-field simulations show that
the modification of the turbulent heat flux by the non-uniform large-scale motions
results in a strong decrease of the critical effective Rayleigh number
required for the formation of the large-scale rolls.
These mean-field simulations have demonstrated that
the spatial distribution of the mean potential temperature
has regions with a positive vertical gradient of the potential temperature
inside the convective roll due to the mean heat flux of the convective rolls.
This study might be useful for understanding the origin
of large-scale rolls observed in atmospheric convective boundary layers
as well as in numerical simulations and laboratory experiments.

\begin{acknowledgments}
I.R. would like to thank the Isaac Newton Institute for Mathematical Sciences,
Cambridge, for support and hospitality during the programme "Anti-diffusive dynamics:
from sub-cellular to astrophysical scales", where the final version of the paper
was completed.
The work of N.K. was partially supported by the Russian Science Foundation (grant 21-72-20067).
The authors benefited from stimulating discussions
of various aspects of turbulent convection with A.~Brandenburg
and P.~J.~K\"{a}pyl\"{a}.
\end{acknowledgments}

\bigskip
\noindent
{\bf DATA AVAILABILITY}
\medskip

The data that support the findings of this study are available from the corresponding author
upon reasonable request.

\medskip
\noindent
{\bf AUTHOR DECLARATIONS}

\noindent
{\bf Conflict of Interest}

The authors have no conflicts to disclose.



\begin{thebibliography}{}

\bibitem {CH61}
S. Chandrasekhar, {\it Hydrodynamic and Hydromagnetic Stability}
(Oxford Univ. Press, Oxford, 1961).

\bibitem{T73}
J. S. Turner, {\em Buoyancy Effects in Fluids}
(Cambridge University Press, Cambridge, 1973).

\bibitem {Z91}
S. S. Zilitinkevich, {\it Turbulent Penetrative
Convection}  (Avebury Technical, Aldershot, 1991).

\bibitem{BPA00}
E. Bodenschatz, W. Pesch, and G. Ahlers,
Recent developments in Rayleigh-B\'{e}nard convection,
Annu. Rev. Fluid Mech. {\bf 32}, 709 (2000).

\bibitem{AGL09}
G. Ahlers, S. Grossmann, D. Lohse,
Heat transfer and large scale dynamics in turbulent Rayleigh-Bénard convection,
Rev. Mod. Phys. {\bf 81}, 503 (2009).

\bibitem{LX10}
D. Lohse, K.-Q. Xia,
Small-scale properties of turbulent Rayleigh-B{\'e}nard convection,
Annu. Rev. Fluid Mech. {\bf 42}, 335 (2010).

\bibitem {ZX10}
Q. Zhou and K.-Q. Xia,
The mixing evolution and geometric properties of a passive scalar field in turbulent Rayleigh–B\'{e}nard convection,
New J. Physics {\bf 12}, 083029 (2010).

\bibitem {MY13} A. S. Monin and A. M. Yaglom, {\em Statistical Fluid
Mechanics}  (Dover, New York, 2013).

\bibitem {LE08} M. Lesieur, {\em Turbulence in Fluids} (Springer, Dordrecht, 2008).

\bibitem {D13} P. A. Davidson, {\em Turbulence in Rotating, Stratified and
Electrically Conducting Fluids} (Cambridge University
Press, Cambridge, 2013).

\bibitem {C14}
E. S. C. Ching, {\it Statistics and Scaling in Turbulent Rayleigh-B\'{e}nard
Convection} (Springer, Singapore, 2014).

\bibitem {RI21}
I. Rogachevskii, {\it Introduction to Turbulent Transport of Particles, Temperature
and Magnetic Fields} (Cambridge University Press, Cambridge, 2021).

\bibitem {KH81}
R. Krishnamurti and L. N. Howard, Large-scale flow generation in
turbulent convection, Proc. Natl.Acad. Sci. USA {\bf 78}, 1981
(1981).

\bibitem {SWL89}
M. Sano, X. Z. Wu and A. Libchaber, Turbulence in helium-gas free
convection, Phys. Rev. A {\bf 40}, 6421 (1989).

\bibitem {CCL96}
S. Ciliberto, S. Cioni and C. Laroche, Large-scale flow properties
of turbulent thermal convection, Phys. Rev. E {\bf 54}, R5901
(1996).

\bibitem {NSS01}
J. J. Niemela, L. Skrbek, K. R. Sreenivasan and R. J. Donnelly,
The wind in confined thermal convection, J. Fluid Mech. {\bf 449}, 169
(2001).

\bibitem {NS03}
J. J. Niemela and K. R. Sreenivasan, Rayleigh-number evolution of
large-scale coherent motion in turbulent convection, Europhys.
Lett. {\bf 62}, 829 (2003).

\bibitem {BKT03}
U. Burr, W. Kinzelbach, A. Tsinober, Is the turbulent wind in
convective flows driven by fluctuations?, Phys. Fluids {\bf 15},
2313 (2003).

\bibitem {XL04}
H. D. Xi, S. Lam and  X. Q. Xia, From laminar plumes to organized
flows: the onset of large-scale circulation in turbulent thermal
convection, J. Fluid Mech. {\bf 503}, 47 (2004).

\bibitem {FA04}
D. Funfschilling and G. Ahlers,
Plume motion and large-scale circulation in a cylindrical Rayleigh-B{\'e}nard cell,
Phys. Rev. Lett. {\bf 92}, 194502 (2004).

\bibitem {BNA05}
E. Brown, A. Nikolaenko and G. Ahlers,
Orientation changes of the
large-scale circulation in turbulent Rayleigh-B{\'e}nard
convection, Phys. Rev. Lett. {\bf 95}, 084503 (2005).

\bibitem {LLT05}
A. Liberzon, B. L\"{u}thi, M. Guala, W. Kinzelbach and A. Tsinober,
Experimental study of the structure of flow regions with negative turbulent
kinetic energy production in confined three-dimensional shear flows with and without
buoyancy,
Phys. Fluids {\bf 17}, 095110 (2005).

\bibitem {EEKR06}
A. Eidelman, T. Elperin, N. Kleeorin, A. Markovich and I.
Rogachevskii,
Hysteresis phenomenon in turbulent convection,
Experim. Fluids {\bf 40}, 723 (2006).

\bibitem {FBA08}
D. Funfschilling, E. Brown and G. Ahlers,
Torsional oscillations of the large-scale circulation in turbulent Rayleigh-B{\'e}nard convection,
J. Fluid Mech {\bf 607}, 119 (2008).

\bibitem {BEKR09}
M. Bukai, A. Eidelman, T. Elperin, N. Kleeorin, I. Rogachevskii and I.
Sapir-Katiraie,
Effect of large-scale coherent structures on turbulent convection,
Phys. Rev. E {\bf 79}, 066302 (2009).

\bibitem {BEKR11}
M. Bukai, A. Eidelman, T. Elperin, N. Kleeorin,
I. Rogachevskii and I. Sapir-Katiraie,
Transition phenomena in unstably stratified turbulent flows,
Phys. Rev. E {\bf 83}, 036302 (2011).

\bibitem {KF84}
J. C. Kaimal and  J. J. Fennigan, {\em Atmospheric Boundary Layer Flows} (Oxford University Press, New York, 1994).

\bibitem {ET85}
D. Etling, Some aspects on helicity in atmospheric flows, Contrib. Atmos. Physics {\bf 58},
88 (1985).

\bibitem {EB93}
D. Etling and R. A. Brown, Roll vortices in the planetary boundary
layer: a review, Boundary-Layer Meteorol. {\bf 65}, 215 (1993).

\bibitem {AZ96}
B. W. Atkinson and J. Wu Zhang,
Mesoscale shallow convection in the
atmosphere, Rev. Geophys. {\bf 34}, 403 (1996).

\bibitem {BR99}
B. Br\"{u}mmer, Roll and cell convection in winter-time arctic cold-air outbreaks, J. Atmosph. Sci.
{\bf  56}, 2613 (1999).

\bibitem {YKH02}
G. S. Young, D. A. R. Kristovich, M. R. Hjelmfelt and R. C. Foster,
Rolls, streets, waves and more,
BAMS, July, ES54 (2002).

\bibitem {W10}
J. C. Wyngaard, {\em Turbulence in the Atmosphere} (Cambridge University Press, 2010).

\bibitem {T17}
F. Tampieri, {\em Turbulence and Dispersion in the Planetary Boundary Layer}
(Springer, 2017).

\bibitem {JB21}
B. Jayaraman and J. G. Brasseur, Transition in atmospheric boundary layer
turbulence structure from neutral to convective, and large-scale rolls, J. Fluid
Mech. {\bf 913}, A42 (2021).

\bibitem {ZK21}
S. Zilitinkevich, E. Kadantsev, I. Repina, E. Mortikov, and A. Glazunov, “Order
out of chaos: Shifting paradigm of convective turbulence,” J. Atmos. Sci. {\bf 78},
3925 (2021).

\bibitem {RKZ22}
I. Rogachevskii, N. Kleeorin, and S. Zilitinkevich,
Energy-and flux-budget theory for surface layers in atmospheric
convective turbulence, Phys. Fluids {\bf 34}, 116602 (2022).

\bibitem {RK24}
I. Rogachevskii and N. Kleeorin,
Semi-organized structures and turbulence in the atmospheric convection,
Phys. Fluids {\bf 36}, 026610 (2024).

\bibitem {HTB03}
T. Hartlep, A. Tilgner and F. H. Busse,
Large-scale structures in Rayleigh-Benard convection at high Rayleigh numbers,
Phys. Rev. Lett. {\bf 91}, 064501 (2003).

\bibitem {PHP04}
A. Parodi, J. von Hardenberg, G. Passoni, A. Provenzale and E. A.
Spiegel, Clustering of plumes in turbulent convection,
Phys. Rev. Lett. {\bf 92}, 194503 (2004).

\bibitem {R06}
F. Rincon, Anisotropy, inhomogeneity and inertial-range scalings in turbulent convection,
J. Fluid Mech. {\bf 563}, 43 (2006).

\bibitem {S08}
J. Schumacher, Lagrangian dispersion and heat transport in convective turbulence,
Phys. Rev. Lett. {\bf 100} 134502 (2008).

\bibitem{CS12}
F. Chill\`{a}, J. Schumacher, New perspectives in turbulent Rayleigh-Bénard convection, Eur. Phys. J. E {\bf 35}, 58 (2012).

\bibitem{AB16}
A. Brandenburg, Stellar mixing length theory with entropy rain,
Astrophys. J. {\bf 832}, 6  (2016).

\bibitem{HZ13}
A. Hellsten and S. Zilitinkevich, Role of convective structures and background
turbulence in the dry convective boundary layer, Boundary-Layer Meteorol.
{\bf 149}, 323 (2013).

\bibitem{KRB17}
P. J. K\"{a}pyl\"{a},  M. Rheinhardt, A. Brandenburg, R. Arlt,  M. J. K\"{a}pyl\"{a}, A. Lagg, N. Olspert, J. Warnecke, Extended subadiabatic layer in simulations of overshooting convection, Astrophys. J. Lett. {\bf 845}, L23 (2017).

\bibitem{KA19}
P. J. K\"{a}pyl\"{a}, Overshooting in simulations of compressible convection,
Astron. Astrophys. {\bf 631}, A122  (2019).

\bibitem{SS20}
J. Schumacher and K. R. Sreenivasan, Colloquium: Unusual dynamics of convection in the Sun,
Rev. Mod. Phys. {\bf 92}, 041001 (2020).

\bibitem{PSS21}
A. Pandey, J. Schumacher and K. R. Sreenivasan, Non-Boussinesq low-Prandtl-number convection with a temperature-dependent thermal diffusivity, Astrophys. J. {\bf 907}, 56 (2021).

\bibitem {EKRZ02}
T. Elperin, N. Kleeorin, I. Rogachevskii and S.S. Zilitinkevich,
Formation of large-scale semi-organized structures in turbulent convection,
Phys. Rev. E {\bf 66}, 066305 (2002).

\bibitem {EKRZ06}
T. Elperin, N. Kleeorin, I. Rogachevskii and S.S. Zilitinkevich,
Tangling turbulence and semi-organized structures in convective boundary layers,
Boundary-Layer Meteorology {\bf 119}, 449-472
(2006).

\bibitem {EGKR06}
T. Elperin, I. Golubev, N. Kleeorin, I. Rogachevskii,
Large-scale instabilities in a nonrotating turbulent convection,
Phys. Fluids {\bf 18}, 126601 (2006).

\bibitem {OKR22}
G. Orian, A. Asulin,  E. Tkachenko, N. Kleeorin, A. Levy, and Rogachevskii I., Large-scale circulations in a shear-free convective turbulence: Mean-field simulations, Phys. Fluids  {\bf  34}, 105121  (2022).

\bibitem {EKRL23}
E. Elmakies, O. Shildkrot, N. Kleeorin, A. Levy and I. Rogachevskii,
Experimental study of turbulent thermal diffusion of particles in an
inhomogeneous forced convective turbulence, Phys. Fluids
{\bf 35}, 095123 (2023).

\bibitem {KKR16}
P. J. K\"{a}pyl\"{a}, A. Brandenburg, N. Kleeorin, M. J. K\"{a}pyl\"{a}, and I. Rogachevskii,
Magnetic flux concentrations from turbulent stratified convection,
Astron. Astrophys. {\bf  683}, A221 (2016).

\bibitem {KA24}
P. J. K\"{a}pyl\"{a}, Convective scale and subadiabatic layers
in simulations of rotating compressible convection,
Astron. Astrophys. {\bf  683}, A221 (2024).

\bibitem {BEKR20}
L. Barel, A. Eidelman, T. Elperin, G. Fleurov, N. Kleeorin, A. Levy, I. Rogachevskii and O. Shildkrot,
Detection of standing internal gravity waves in experiments with convection over a wavy heated wall,
Phys. Fluids {\bf 32}, 095105 (2020).

\bibitem {TSW17}
S. Toppaladoddi, S. Succi, and J. S. Wettlaufer,
Roughness as a route to the ultimate regime of thermal convection,
Phys. Rev. Lett. {\bf 118}, 074503 (2017).

\end{thebibliography}
\end{document}